\documentclass[a4paper,11pt]{article}
\pdfoutput=1
\usepackage{jheppub}
\usepackage{lineno}
\usepackage{multirow}
\usepackage{makecell} 
\usepackage{booktabs}
\usepackage{amsmath}
\usepackage{bbm}
\usepackage{textcomp} 

\allowdisplaybreaks[4]

\newcommand{\SKLP}{State Key Laboratory of Particle Detection and Electronics, University of Science and Technology of China, Hefei 230026, Anhui, People’s Republic of China}
\newcommand{\USTC}{Department of Modern Physics, University of Science and Technology of China, Hefei 230026, Anhui, People’s Republic of China}

\arxivnumber{2401.01323}

\title{\boldmath Mixed QCD-EW corrections to $W$-pair production at electron-positron colliders}

\author[a,b]{Zhe Li,}
\author[a,b,c]{Ren-You Zhang,}
\author[a,b]{Shu-Xiang Li,}
\author[a,b]{Xiao-Feng Wang,}
\author[a,b]{Wen-Jie He,}
\author[a,b]{Liang Han,}
\author[a,b]{Yi Jiang,}
\author[d]{Qing-hai Wang\,}

\emailAdd{brucelee@mail.ustc.edu.cn}
\emailAdd{zhangry@ustc.edu.cn}
\emailAdd{lishux@mail.ustc.edu.cn}
\emailAdd{xf0914@mail.ustc.edu.cn}
\emailAdd{wjbox@mail.ustc.edu.cn}
\emailAdd{hanl@ustc.edu.cn}
\emailAdd{jiangyi@ustc.edu.cn}
\emailAdd{qhwang@nus.edu.sg}

\affiliation[a]{\SKLP}
\affiliation[b]{\USTC}
\affiliation[c]{Anhui Center for Fundamental Sciences in Theoretical Physics, University of Science and Technology of China, Hefei 230026, Anhui, People’s Republic of China}
\affiliation[d]{Department of Physics, National University of Singapore, Singapore 117551, Singapore}

\abstract{
The discrepancy between the CDF measurement and the Standard Model theoretical prediction for the $W$-boson mass underscores the importance of conducting high-precision studies on the $W$ boson, which is one of the predominant objectives of proposed future $e^+e^-$ colliders. We investigate in detail the production of $W$-boson pairs at $e^+e^-$ colliders, and compute the next-to-next-to-leading order mixed QCD-EW corrections to both the integrated cross section and various kinematic distributions. By employing the method of differential equations, we analytically calculate the two-loop master integrals for the mixed QCD-EW virtual corrections to  $e^+e^- \rightarrow W^+W^-$. Utilizing the Magnus transformation, we derive a set of canonical master integrals for each integral family. This canonical basis satisfies a system of differential equations in which the dependence on the dimensional regulator is linearly factorized from the kinematics. We then express all these canonical master integrals as Taylor series in $\epsilon$ up to $\epsilon^4$, with coefficients articulated in terms of Goncharov polylogarithms up to weight four. Upon applying our analytic expressions of these master integrals to the phenomenological analysis of $W$-pair production, we observe that the $\mathcal{O}(\alpha\alpha_s)$ corrections are significantly impactful in the $\alpha(0)$ scheme, particularly in certain phase-space regions. However, these mixed QCD-EW corrections can be heavily suppressed by adopting the $G_{\mu}$ scheme.
}


\begin{document}
\maketitle
\flushbottom

\section{Introduction}
\label{sec:1}
\par
The discovery of the Higgs boson \cite{ATLAS:2012yve,CMS:2012qbp} at the CERN Large Hadron Collider (LHC) in 2012 marked a pivotal moment in the field of elementary particle physics, validating the last missing piece of the Standard Model (SM). However, a recent high-precision measurement of the $W$-boson mass by the CDF collaboration \cite{CDF:2022hxs},
\begin{equation}
m_{\scriptscriptstyle{W}}^{\scriptscriptstyle{\text{CDF}}} = 80.4335 \pm 0.0094~ \mathrm{GeV}\,,
\end{equation}
reveals a significant $7 \sigma$ deviation from the SM prediction, challenging the internal consistency of the SM. Addressing this discrepancy necessitates a deep and thorough understanding of the SM, especially the gauge structure of the electroweak (EW) sector. Consequently, the pursuit of high-precision experimental measurements and refined theoretical studies within the SM framework remains a critical goal in both current and future high-energy physics research.

\par
The production of $W$-boson pairs at $e^+e^-$ colliders offers a direct avenue for measuring the $W$-boson mass, since the production cross section around the threshold is highly sensitive to the $W$-boson mass. Furthermore, this process serves as an ideal platform for investigating the electroweak symmetry breaking mechanism, as it directly unveils the intricate structure of triple gauge-boson interactions \cite{Bilenky:1993ms,OPAL:2003xqq}. Remarkably, the total cross section of $W$-pair production has been measured with an impressive accuracy of approximately $1\%$, and the determination of the $W$-boson mass has reached a precision of $33~ \mathrm{MeV}$, achieved through a combination of direct reconstruction and threshold measurements at LEP2 \cite{ALEPH:2013dgf}. Future advancements in precision measurements of the $W$-boson mass and the total cross section of $W$-pair production are anticipated. These endeavors are supported by proposals from next-generation high-luminosity $e^+e^-$ collider projects, including the International Linear Collider (ILC) \cite{Behnke:2013xla,ILC:2013jhg,Bambade:2019fyw}, the Circular Electron-Positron Collider (CEPC) \cite{CEPCStudyGroup:2018rmc,CEPCStudyGroup:2018ghi} and the Future Circular Collider (FCC-ee) \cite{FCC:2018byv,FCC:2018evy}. Notably, these initiatives aim to measure the $W$-boson mass with extraordinary precision, achieving accuracies of just a few $\mathrm{MeV}$ at ILC \cite{ILC:2013jhg}, $1~ \mathrm{MeV}$ at CEPC \cite{CEPCStudyGroup:2018ghi} and $0.5~ \mathrm{MeV}$ at FCC-ee \cite{FCC:2018evy}, surpassing the precision of the CDF measurement.

\par
In anticipation of forthcoming high-precision experimental measurements, it is crucial to achieve an extremely fine level of control over the theoretical prediction for the $W$-pair production cross section, aiming for permille (or even sub-permille) precision. The process of $e^+e^- \rightarrow W^+W^-$ has been extensively studied at LEP, specifically focusing on the measurement of $W$ helicity and the effects of beam polarization \cite{Blondel:1987pm}. The complete next-to-leading order (NLO) electroweak (EW) corrections to $e^+e^- \rightarrow W^+W^-$, comprising EW one-loop virtual corrections, real-photon radiation corrections and leading-logarithmic (LL) initial-state QED corrections, have been calculated over the past few decades \cite{Alles:1976qv,Hagiwara:1986vm,Lemoine:1979pm,Philippe:1981up,Bohm:1987ck,Beenakker:1990sf,Beenakker:1991jk,Fleischer:1988kj,Kolodziej:1991pk,Fleischer:1991nw,Beenakker:1994vn,Zerwas:1991rrh}. For more comprehensive overviews, please refer to refs. \cite{Denner:1991kt,Beenakker:1996kt}.

\par
Despite the remarkable precision achieved by NLO EW theoretical predictions, often attaining an accuracy of a few percent or even permille, it is anticipated that the precision of experimental measurements at forthcoming facilities, such as ILC, CEPC, and FCC-ee, will surpass this level of accuracy. To align with the expected permille accuracy of cross section measurements at future lepton colliders, it is essential to delve into higher-order corrections in theoretical predictions. The next-to-next-to-leading order (NNLO) corrections to EW processes include both pure EW corrections and mixed QCD-EW corrections. Calculating the EW $\mathcal{O}(\alpha^2)$ corrections presents a significant challenge due to the tremendous number of two-loop Feynman diagrams involved in virtual corrections. Conversely, the mixed NNLO QCD-EW corrections are more tractable, and may possess a greater magnitude. In light of these considerations, this paper focuses on a comprehensive analysis of the  $\mathcal{O}(\alpha\alpha_s)$ corrections to the $W$-pair production cross section at lepton colliders, representing the most refined and precise theoretical prediction to date.

\par
The mixed QCD-EW $\mathcal{O}(\alpha\alpha_s)$ corrections to $e^+e^- \rightarrow W^+W^-$ arise from the interference between the leading order (LO) and $\text{QCD} \otimes \text{EW}$ NNLO amplitudes. These corrections can be categorized into two types: vertex corrections and self-energy corrections. Notably, the NNLO QCD-EW corrections to $e\nu_{e}W$, $ee\gamma$ and $eeZ$ vertices are exclusively contributed by their respective $\mathcal{O}(\alpha\alpha_s)$ counterterms. In this paper, we undertake an analytic calculation of the two-loop Feynman integrals present in the $\text{QCD} \otimes \text{EW}$ NNLO amplitude. We then utilize these analytic results to derive the NNLO QCD-EW corrected integrated cross section and various kinematic distributions. Of particular significance is our thorough analytic treatment of the two-loop triangle master integrals (MIs) for mixed QCD-EW triple gauge-boson vertex corrections, which stem from a gluon-dressed quark loop with two distinct massive flavors. It is pertinent to mention that the MIs with massless flavors and with only one massive flavor in the quark loop have been extensively studied in refs. \cite{Usyukina:1994iw,Birthwright:2004kk,Chavez:2012kn,DiVita:2017xlr,Ma:2021cxg}.

\par
The rest of this paper is organized as follows. In section \ref{sec:2}, we begin by establishing our notations for the calculation of $e^+e^- \rightarrow W^+W^-$, and proceed to delve into the details of the NLO EW corrections and the NNLO QCD-EW corrections. Section \ref{sec:3} is dedicated to the analytic calculation of the MIs essential for the mixed QCD-EW two-loop corrections to the triple gauge-boson couplings (TGCs) $VWW~ (V = \gamma, Z)$. We elaborate on the construction and the solution of the canonical differential equations, as well as the analytic continuation of the MIs. Utilizing the analytic expressions of the MIs derived in section \ref{sec:3}, we compute the production cross section and certain kinematic distributions for $e^+e^- \rightarrow W^+W^-$ at the QCD-EW NNLO in both the $\alpha(0)$ and $G_{\mu}$ schemes. The numerical results and a comprehensive discussion are provided in section \ref{sec:4}. Finally, a brief summary is given in section \ref{sec:5}.

\section{Descriptions of perturbative calculations}
\label{sec:2}
\par
In this paper, we focus on the calculation of the mixed QCD-EW corrections to the scattering process
\begin{equation}
e^+(p_1, \lambda_1) + e^-(p_2, \lambda_2) \rightarrow W^+(p_3, \lambda_3) + W^-(p_4, \lambda_4)\,, 
\end{equation}
where $p_1^2 = p_2^2 = 0$, $p_3^2 = p_4^2 = m_{\scriptscriptstyle{W}}^2$, and the electron mass is consistently neglected wherever feasible. Here, $\lambda_{1, 2}$ denote the helicities of the initial-state positron and electron, respectively, and $\lambda_{3, 4}$ represent the polarizations of the final-state $W^{\pm}$ bosons. The Mandelstam invariants for this $2 \rightarrow 2$ scattering process are defined as
\begin{align}
s &= (p_1 + p_2)^2  = 4\, E^2\,,
\nonumber \\
t &= (p_1 - p_3)^2 = m_{\scriptscriptstyle{W}}^2 - 2\, E^2\, (1 - \beta \cos\theta)\,,
\\
u &= (p_1 - p_4)^2 = m_{\scriptscriptstyle{W}}^2 - 2\, E^2\, (1 + \beta \cos\theta)\,,
\nonumber
\end{align}
with $E$ representing the beam energy, $\theta$ the scattering angle between $e^+$ and $W^+$, and $\beta=\sqrt{1-m_{\scriptscriptstyle{W}}^2/E^2}$ the velocity of the $W^{\pm}$ bosons in the center-of-mass (CM) frame. The LO unpolarized differential cross section in the CM frame is given by 
\begin{equation}
\frac{\mathrm{d} \sigma_{\scriptscriptstyle{\text{LO}}}}{\mathrm{d} \Omega}
=
\frac{\beta}{64 \pi^2 s}\, \frac{1}{4} \sum_{\lambda_{1, \ldots, 4}} \big| \mathcal{M}_0(s, t, \lambda_{1, \ldots, 4})\big|^2\,,
\end{equation}
where $\mathcal{M}_0$ is the lowest-order amplitude for $e^+e^- \rightarrow W^+W^-$.

\subsection{NLO EW corrections}
\label{sec:2.1}
\par
There are two dominant channels for $W$-pair production at electron-positron colliders: the $t$-channel via $\nu_e$ exchange, exclusively contributed by left-handed electrons, and the $s$-channel via $\gamma$ or $Z$ exchange, involving both left- and right-handed electrons. Notably, the contribution from Higgs exchange is entirely disregarded due to the exceedingly small Yukawa coupling involved.

\par
The fundamental characteristics of $W$-pair production are governed by the Born cross section. Near the threshold region ($\beta \ll 1$), the unpolarized integrated cross section in the Born approximation is expressed as \cite{Denner:1991kt}
\begin{equation}
\sigma_{\text{Born}}
=
\frac{\pi \alpha^2}{s}\, \frac{1}{\sin^2\theta_{\scriptscriptstyle{\text{W}}}}\, \beta + \mathcal{O}(\beta^3)\,.
\end{equation}
The leading term, proportional to $\beta$, originates exclusively from the $t$-channel, resulting in the threshold behavior of the cross section for $W$-boson pair production in $e^+e^-$ annihilation. In contrast, contributions from the $s$-channel and $s$-$t$ interference, which are proportional to $\beta^3$, become negligible near the threshold compared to the $t$-channel contribution. In the high-energy region, the effects of triple gauge-boson interactions from the $s$-channel become more pronounced, particularly at large scattering angles. To improve the sensitivity to TGCs, one could utilize right-handed polarized electrons to filter out the $t$-channel contribution. For more comprehensive analysis, please refer to refs. \cite{Alles:1976qv,Hagiwara:1986vm,Denner:1991kt,Beenakker:1994vn,Beenakker:1996kt}.

\par
The complete $\mathcal{O}(\alpha)$ corrections consist of two components: the virtual one-loop correction and the real photon radiation correction. Moreover, it is essential to incorporate the initial-state radiation (ISR) effect, which can be implemented using the LL approximation. The $\mathcal{O}(\alpha)$ virtual correction to the differential cross section in the CM frame is given by
\begin{equation}
\frac{\mathrm{d} \sigma_{\text{virtual}}}{\mathrm{d} \Omega}
=
\frac{\beta}{64 \pi^2 s}\, \frac{1}{4} \sum_{\lambda_{1, \ldots, 4}}
2\, \mathrm{Re}\,
\big[\,
\mathcal{M}_0^{\ast}(s, t, \lambda_{1, \ldots, 4})\, \mathcal{M}_{\text{1-loop}}(s, t, \lambda_{1, \ldots, 4})
\,\big]\,,
\end{equation}
where $\mathcal{M}_{\text{1-loop}}$ represents the one-loop level amplitude. In our NLO calculation, we adopt the 't Hooft-Feynman gauge and the on-shell (OS) renormalization scheme \cite{Denner:1991kt,Denner:2019vbn}. The ultraviolet (UV) divergences arising from the loop amplitude are regularized by dimensional regularization (DR) in $d= 4-2\, \epsilon$ dimensions \cite{tHooft:1972tcz,Bollini:1972ui}, and are cancelled after performing the renormalization procedure. The infrared (IR) divergences induced by virtual photon loops are regularized by introducing an infinitesimal fictitious photon mass. The inclusion of real photon radiation ensures the cancellation of these IR divergences. To isolate the IR singularities arising from real photon radiation, we employ the two cutoff phase space slicing method \cite{Harris:2001sx}, introducing two arbitrary cutoffs, $\delta_s$ and $\delta_c$, to partition the phase space of real photon emission into soft (S), hard-collinear (HC) and hard-noncollinear ($\text{H}\overline{\text{C}}$) regions. The $\mathcal{O}(\alpha)$ real photon radiation correction is thus decomposed as
\begin{equation}
\sigma_{\text{real}}
=
\sigma_{\scriptscriptstyle{\text{S}}}(\delta_s)
+
\sigma_{\scriptscriptstyle{\text{HC}}}(\delta_s, \delta_c)
+
\sigma_{\scriptscriptstyle{\text{H}}\overline{\scriptscriptstyle{\text{C}}}}(\delta_s, \delta_c)\,.
\end{equation}
We confirmed the cutoff independence of the real photon radiation correction within the range $\delta_s = 50\, \delta_c \in [10^{-6},10^{-3}]$. All results have been cross-verified using the Catani-Seymour dipole subtraction scheme \cite{Catani:1996vz,Catani:2002hc,Dittmaier:1999mb,Dittmaier:2008md}.

\par
The LL QED correction due to ISR can be formulated as a convolution of the Born-level cross section with structure functions \cite{Beenakker:1996kt,Denner:2000bj},
\begin{equation}
\sigma_{\scriptscriptstyle{\text{ISR-LL}}}
=
\int_0^1 \mathrm{d}x_1 \mathrm{d}x_2\,
\Gamma_{ee}^{\scriptscriptstyle{\text{LL}}}(x_1, Q^2)\,
\Gamma_{ee}^{\scriptscriptstyle{\text{LL}}}(x_2, Q^2)
\int \mathrm{d}\sigma_{\scriptscriptstyle{\text{LO}}}(x_1 p_1, x_2 p_2)\,,
\end{equation}
where $x_1$ and $x_2$ denote the fractions of the longitudinal momenta carried by the initial-state leptons after photon radiation. The typical scale of the hard scattering process, $Q^2$, is chosen as $s$. The LL structure function $\Gamma_{ee}^{\scriptscriptstyle{\text{LL}}}(x, Q^2)$ is given explicitly up to $\mathcal{O}(\alpha^3)$ in refs. \cite{Beenakker:1996kt,Denner:2000bj}. To avoid double counting, the Born-level cross section and the one-photon emission correction must be subtracted from the ISR contribution. Consequently, the higher-order initial-state radiation (h.o.ISR) correction can be expressed as
\begin{equation}
\sigma_{\scriptscriptstyle{\text{h.o.ISR}}}
=
\sigma_{\scriptscriptstyle{\text{ISR-LL}}} - \sigma_{\scriptscriptstyle{\text{LL,sub}}}\,,
\end{equation}
where the subtraction term $\sigma_{\scriptscriptstyle{\text{LL,sub}}}$ is given by
\begin{align}
\sigma_{\scriptscriptstyle{\text{LL,sub}}}
=
\int_0^1 \mathrm{d}x_1 \mathrm{d}x_2\,
&
\Big[\,
\delta(1 - x_1)\, \delta(1 - x_2)
\\
&
+ \Gamma_{ee}^{\scriptscriptstyle{\text{LL}}}(x_1, Q^2)\, \delta(1 - x_2)
+ \Gamma_{ee}^{\scriptscriptstyle{\text{LL}}}(x_2, Q^2)\, \delta(1 - x_1)
\,\Big]
\int \mathrm{d}\sigma_{\scriptscriptstyle{\text{LO}}}(x_1 p_1, x_2 p_2)\,.
\nonumber
\end{align}
Ultimately, the full EW correction is defined as the collective sum of the virtual one-loop correction, the real photon radiation correction and the h.o.ISR correction,
\begin{equation}
\Delta \sigma_{\scriptscriptstyle{\text{EW}}}
=
\sigma_{\text{virtual}} + \sigma_{\text{real}} + \sigma_{\scriptscriptstyle{\text{h.o.ISR}}}\,.
\end{equation}

\par
Particular attention must be directed towards the electric charge renormalization. In the $\alpha(0)$ scheme, the fine structure constant is defined from the $ee\gamma$ coupling for on-shell external particles in the Thomson limit. The electric charge renormalization constant in this scheme is given by
\begin{equation}
\delta Z_{e,0}
=
-\, \frac{1}{2}\, \delta Z_{\gamma\gamma}
- \frac{1}{2} \tan\theta_{\scriptscriptstyle{\text{W}}}\, \delta Z_{{\scriptscriptstyle{Z}}\gamma}
=
\frac{1}{2}\, \Pi^{\gamma\gamma}(0)
- \tan\theta_{\scriptscriptstyle{\text{W}}}\, \frac{\Sigma_{\text{T}}^{\gamma\scriptscriptstyle{Z}}(0)}{m_{\scriptscriptstyle{Z}}^2}\,,
\end{equation}
which contains mass-singular terms $\log (m^2_{f}/Q^2)~ (f = e, \mu, \tau, u, d, c, s, b)$. Notably, for QED couplings with external photons, the large logarithms arising from the electric charge renormalization constant are precisely cancelled by those from the wave-function renormalization constant of the external photon in the $\alpha(0)$ scheme. For other EW couplings, the mass-singular terms of $\delta Z_{e,0}$ can be absorbed into the running fine structure constant by using the $G_\mu$ scheme, wherein the fine structure constant is derived from the Fermi constant $G_{\mu}$ through the following relation:
\begin{equation}
\label{eq:alphaGmu}
\alpha_{\scriptscriptstyle{G_{\mu}}}
=
\frac{\sqrt{2}\, G_{\mu}\, m_{\scriptscriptstyle{W}}^2}{\pi}
\Big(1 - \frac{m_{\scriptscriptstyle{W}}^2}{m_{\scriptscriptstyle{Z}}^2} \Big)\,.
\end{equation}
The electric charge renormalization constant in the $G_{\mu}$ scheme is then modified to
\begin{equation}
\delta Z_{e,\scriptscriptstyle{G_{\mu}}} = \delta Z_{e,0} - \frac{1}{2}\, \Delta r\,,
\end{equation}
where the subtraction term $\Delta r$ comprises the higher-order corrections to muon decay, given as \cite{Sirlin:1980nh,Dittmaier:2015rxo}
\begin{equation}
\label{eq:Dr}
\Delta r
=
\Pi^{\gamma\gamma}(0)
- 2\, \frac{\delta \sin\theta_{\scriptscriptstyle{\text{W}}}}{\sin\theta_{\scriptscriptstyle{\text{W}}}}
+ 2 \cot\theta_{\scriptscriptstyle{\text{W}}}\, \frac{\Sigma_{\text{T}}^{\gamma\scriptscriptstyle{Z}}(0)}{m_{\scriptscriptstyle{Z}}^2}
+ \frac{\Sigma_{\text{T}}^{\vphantom{\gamma}\scriptscriptstyle{WW}}(0) 
- \mathrm{Re}\, \Sigma_{\text{T}}^{\vphantom{\gamma}\scriptscriptstyle{WW}}(m_{\scriptscriptstyle{W}}^2)}{m_{\scriptscriptstyle{W}}^2}
+ \delta r\,,
\end{equation}
and the $\mathcal{O}(\alpha)$ contribution to the finite remainder $\delta r$ is given by
\begin{equation}
\delta r
=
\frac{\alpha(0)}{4\pi \sin^2\theta_{\scriptscriptstyle{\text{W}}}}
\Big(
6 + \frac{7 - 4 \sin^2\theta_{\scriptscriptstyle{\text{W}}}}{2 \sin^2\theta_{\scriptscriptstyle{\text{W}}}}\,
\log \cos^2\theta_{\scriptscriptstyle{\text{W}}}
\Big)\,.
\end{equation}
For comparison purposes, our calculations are performed in both the $\alpha(0)$ scheme and the $G_{\mu}$ scheme.

\par
To compute the EW corrections to $e^+e^- \rightarrow W^+W^-$, we use our modified \texttt{FormCalc} and \texttt{LoopTools} packages \cite{Hahn:1998yk,vanOldenborgh:1990yc}. When comparing our integrated cross sections to those in refs. \cite{Denner:1991kt,Beenakker:1996kt}, we observe a remarkable agreement, surpassing even the permille level.

\subsection{NNLO mixed QCD-EW corrections}
\label{sec:2.2}
\par
\begin{figure}[htbp]
\centering
\includegraphics[width = 1.0\textwidth]{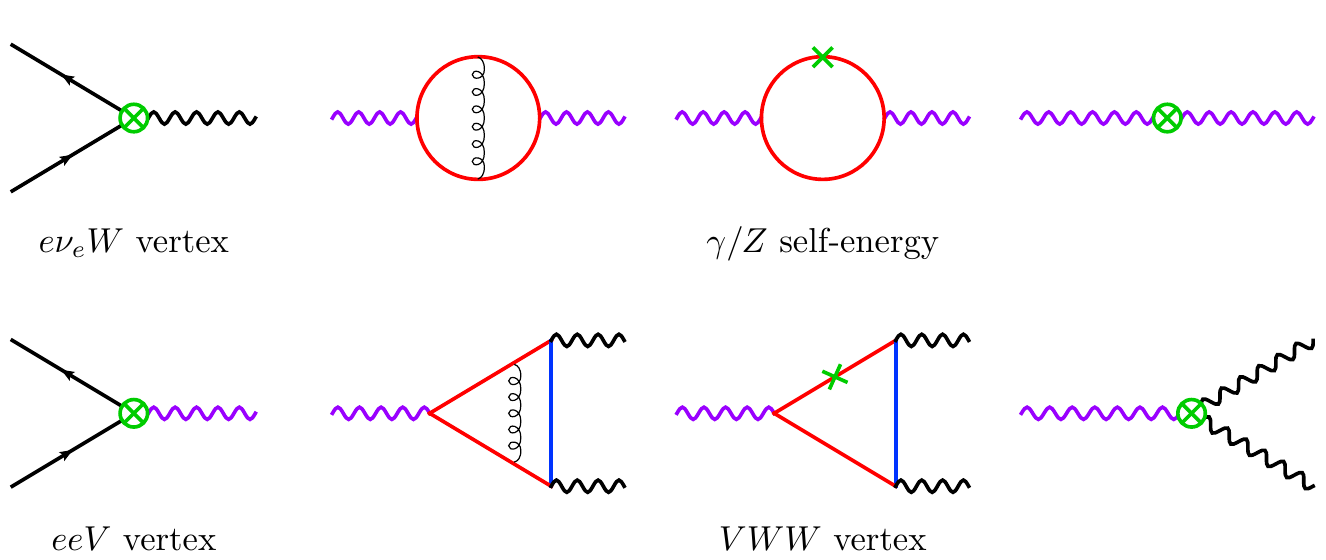} 
\caption{Representative Feynman diagrams for the NNLO mixed QCD-EW corrections to $e^+e^- \rightarrow W^+W^-$. The green crosses symbolize the quark mass counterterm at $\mathcal{O}(\alpha_s)$, whereas the green circled crosses signify the vertex or self-energy counterterm at $\mathcal{O}(\alpha\alpha_s)$.}
\label{fig1}
\end{figure}

The mixed QCD-EW $\mathcal{O}(\alpha\alpha_s)$ corrections to $e^+e^- \rightarrow W^+W^-$ encompass the $e\nu_eW$ vertex correction, $eeV$ vertex correction, $\gamma/Z$ self-energy correction and $VWW$ vertex correction. Representative Feynman diagrams for these $\mathcal{O}(\alpha\alpha_s)$ corrections are illustrated in figure \ref{fig1}. The first two types of corrections originate solely from the $\mathcal{O}(\alpha\alpha_s)$ counterterms, rendering them UV-finite. However, the latter two types include corrections from two-loop diagrams where a gluon is attached to each one-loop quark line in all feasible manners. These corrections also incorporate one-loop diagrams with insertions of $\mathcal{O}(\alpha_s)$ quark mass counterterms,\footnote{By virtue of the QED-like Ward identity,  the vertex and fermion wave-function counterterms exactly cancel each other, leaving contributions solely from quark mass renormalization.} as well as $\mathcal{O}(\alpha\alpha_s)$ vertex and self-energy counterterms. Thus, the $\mathcal{O}(\alpha\alpha_s)$ correction to the amplitude can be decomposed as
\begin{equation}
\mathcal{M}_{\text{2-loop}}
=
\mathcal{M}_{e\nu_e\scriptscriptstyle{W}}
+
\mathcal{M}_{ee\scriptscriptstyle{V}}
+
\mathcal{M}_{\gamma/\scriptscriptstyle{Z\text{-SE}}}
+
\mathcal{M}_{\scriptscriptstyle{VWW}}\,,
\end{equation}
and the corresponding mixed QCD-EW $\mathcal{O}(\alpha\alpha_s)$ correction to the differential cross section in the CM frame is expressed as
\begin{equation}
\frac{\mathrm{d} \sigma_{\scriptscriptstyle{\text{QCD-EW}}}}{\mathrm{d} \Omega}
=
\frac{\beta}{64 \pi^2 s}\, \frac{1}{4} \sum_{\lambda_{1, \ldots, 4}}
2\, \mathrm{Re}\,
\big[\,
\mathcal{M}_0^{\ast}(s, t, \lambda_{1, \ldots, 4})\, \mathcal{M}_{\text{2-loop}}(s, t, \lambda_{1, \ldots, 4})
\,\big]\,.
\end{equation}
The counterterm contributions to the mixed QCD-EW correction consist of the $\mathcal{O}(\alpha_s)$ quark mass counterterm and the $\mathcal{O}(\alpha\alpha_s)$ vertex and self-energy counterterms, involving the quark mass renormalization constant as well as the gauge-boson mass and wave-function renormalization constants. The $\mathcal{O}(\alpha_s)$ quark mass renormalization constant in the OS scheme is given by \cite{Bernreuther:2004ih}
\begin{equation}
\frac{\delta m_q}{m_q}
=
-\, \frac{\alpha_s}{2\pi}\, C(\epsilon)\, \Big(\frac{\mu^2}{m_{q}^2}\Big)^\epsilon\,
\frac{C_F}{2}\, \frac{3 - 2\, \epsilon}{\epsilon\, (1 - 2\, \epsilon)}\,,
\end{equation}
where $C(\epsilon) = (4\pi)^{\epsilon}\, \Gamma(1 + \epsilon)$, $C_F = 4/3$ and $\mu$ is the renormalization scale. The explicit expressions for the $\mathcal{O}(\alpha\alpha_s)$ gauge-boson mass and wave-function renormalization constants are derived from the corresponding NLO EW constants \cite{Denner:1991kt} by replacing the one-loop gauge-boson self-energies with their two-loop $\mathcal{O}(\alpha\alpha_s)$ counterparts, as documented in refs. \cite{Dittmaier:2015rxo,Chang:1981qq,Djouadi:1987gn,Djouadi:1987di,Kniehl:1988ie,Kniehl:1989yc,Djouadi:1993ss}. Furthermore, in the $G_{\mu}$ scheme, the contributions to $\Delta r$ at $\mathcal{O}(\alpha\alpha_s)$ are streamlined due to the absence of the finite remainder $\delta r$ and the vanishing self-energy $\Sigma_{\text{T}}^{\gamma\scriptscriptstyle{Z}}(0)$ at the QCD-EW NNLO. Once all contributions at $\mathcal{O}(\alpha\alpha_s)$ are considered, all UV and IR divergences are precisely cancelled, ensuring a consistent theoretical framework.

\par
In our computational journey of NNLO mixed QCD-EW corrections, we employ the \texttt{FeynArts} package \cite{Hahn:2000kx} to generate Feynman diagrams and their corresponding amplitudes. These amplitudes, after further processing with \texttt{FeynCalc} \cite{Mertig:1990an,Shtabovenko:2020gxv}, are expressed as linear combinations of a significant number of scalar Feynman integrals, which can be categorized into various families. Scalar Feynman integrals within the same family are interrelated and can be systematically reduced to a set of irreducible MIs via integration-by-parts (IBP) identities \cite{Tkachov:1981wb,Chetyrkin:1981qh}. The IBP reduction can be facilitated by utilizing publicly available packages, such as \texttt{Kira} \cite{Maierhofer:2017gsa,Klappert:2020nbg}, \texttt{FIRE} \cite{Smirnov:2019qkx}, \texttt{LiteRed} \cite{Lee:2012cn,Lee:2013mka}, \texttt{FiniteFlow} \cite{Peraro:2019svx}, \texttt{NeatIBP} \cite{Wu:2023upw} and \texttt{Blade} \cite{Guan:2024byi}. In this work, we employ \texttt{Kira} to conduct IBP reduction, wherein the Laporta algorithm \cite{Laporta:2000dsw} is implemented for solving IBP identities. Subsequently, we derive the two-loop MIs for the NNLO mixed QCD-EW corrections to $e^+e^- \rightarrow W^+W^-$, which is the focal point of our calculation. While the two-loop triangle MIs with massless propagators and the self-energy MIs are documented in refs. \cite{Djouadi:1993ss,Usyukina:1994iw}, the two-loop triangle MIs with two distinct massive quarks in the loops remain elusive, presenting significant challenges due to the multiple mass scales involved. We successfully achieved analytic expressions for these two-loop MIs by using the canonical differential equation method \cite{Henn:2013pwa,Henn:2014qga}. Additionally, to ensure the accuracy and reliability of our results, all MIs have been cross-verified with high precision using \texttt{pySecDec} \cite{Borowka:2017idc,Borowka:2018goh} and \texttt{AMFlow} \cite{Liu:2022chg}. Detailed discussions and further elaborations of our analytic calculation for the massive two-loop triangle MIs involved in the $\mathcal{O}(\alpha\alpha_s)$ corrections to $e^+e^- \rightarrow W^+W^-$ are provided in section \ref{sec:3}.

\section{Canonical differential equations}
\label{sec:3}
\par
In this section, we begin by establishing the notations and conventions essential for the calculation of two-loop MIs. Subsequently, we provide a concise overview of the canonical differential equation method. Following this, we delve into the construction and solution of the canonical system, specifically tailored to the $e^+e^- \rightarrow W^+W^-$ process. Finally, we discuss the analytic continuation of the canonical MIs, detailing their extension across various kinematic regions.

\par
\begin{figure}[htbp]
\centering
\includegraphics[width = 0.9\textwidth]{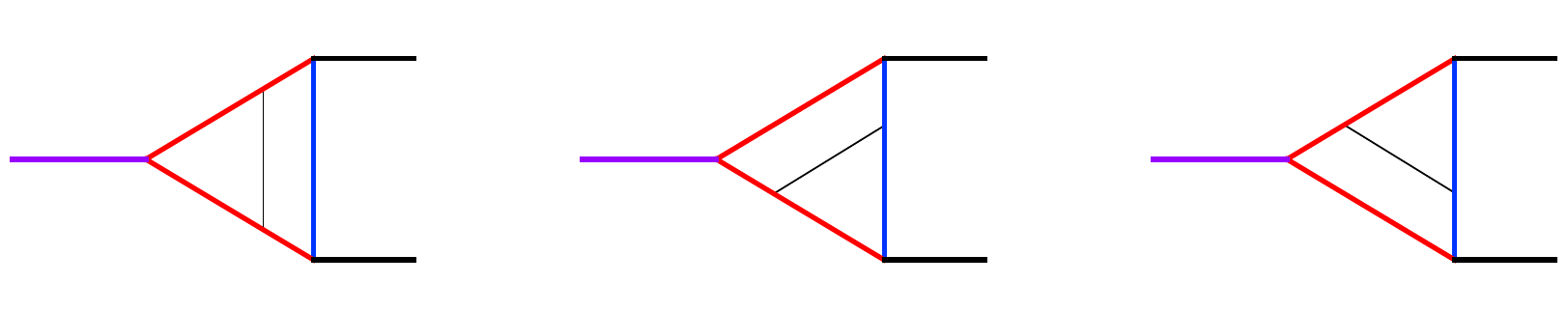} 
\caption{Three top-level topologies of the integral family $\mathcal{F}$. The thin black lines denote massless propagators, while the thick red and blue lines represent top- and bottom-quark propagators, respectively. The thick purple lines indicate an external off-shell leg with momentum squared $s$, and the thick black external lines signify the on-shell $W$ bosons.}
\label{fig2}
\end{figure}

Our primary focus centers on the analytic calculation of the massive two-loop MIs arising from the mixed QCD-EW corrections to $VWW$ vertex,
\begin{equation}
V^{\ast}(p_3 + p_4) \rightarrow W^+(p_3) + W^-(p_4)\,, 
\end{equation}
where $V^{\ast}$ represents an off-shell neutral gauge boson and both $W$ bosons are on-shell. In this paper, the dimensionally regularized two-loop three-point scalar Feynman integrals are defined as
\begin{equation}                                  
F(\alpha_1,\ldots, \alpha_7) = \int \mathcal{D}^d l_1 \mathcal{D}^d l_2\, \frac{1}{D_1^{\alpha_1} \ldots D_7^{\alpha_7}}\,,
\end{equation}
where the integration measure is conveniently chosen as
\begin{equation}
\label{eq:measure}
\mathcal{D}^d l_i
=
\frac{d^d l_i}{(2 \pi)^d}
\left( \frac{i S_\epsilon}{16 \pi^2} \right)^{-1}
\big( m_t^2 \big)^{\epsilon}
\quad
\text{with}
\quad
S_{\epsilon}
=
(4\pi)^{\epsilon}\, \Gamma(1+\epsilon)\,.
\end{equation}
The massive two-loop Feynman diagrams for the $\mathcal{O}(\alpha\alpha_s)$ corrections to the $VWW$ vertex are categorized into six top-level topologies. Three of these, depicted in figure \ref{fig2}, belong to the integral family $\mathcal{F}$, identified by the following set of propagators,
\begin{align}
\label{eq:Ffamily}
&
D_1 = (l_1 + p_3)^2 - m_t^2
&\quad&
D_3 = l_1^2 - m_b^2
&\quad&
D_5 = (l_1 - p_4)^2 - m_t^2
&\quad&
D_7 = (l_1 - l_2)^2
\quad~
\nonumber \\
&
D_2 = (l_2 + p_3)^2 - m_t^2
&\quad&
D_4 = l_2^2 - m_b^2
&\quad&
D_6 = (l_2 - p_4)^2 - m_t^2
\end{align}
The other three top-level topologies, derived from figure \ref{fig2} by exchanging the top and bottom quarks in the loops, belong to the family $\mathcal{F}^{\ast} = \mathcal{F}|_{m_t \leftrightarrow m_b}$.

\subsection{Canonical system}
\label{sec:3.1}
\par
In general, the vector of MIs, denoted as $\mathbf{F}$, satisfies a system of differential equations,
\begin{equation}
d \mathbf{F}(\vec{x}, \epsilon) = d \text{A}(\vec{x}, \epsilon)\, \mathbf{F}(\vec{x}, \epsilon)\,,
\end{equation}
where the coefficient matrix $\text{A}(\vec{x}, \epsilon)$ depends on both the kinematic variable $\vec{x}$ and the dimensional regulator $\epsilon$. It is important to note that selecting a suitable set of MIs, often referred to as a canonical basis, can significantly streamline the calculation of the differential system \cite{Henn:2013pwa,Henn:2014qga}. By employing the Magnus exponential method \cite{magnus1954exponential,blanes2009magnus,Argeri:2014qva}, we can determine the transformation that maps the general basis $\mathbf{F}$ to the canonical basis $\mathbf{I}$, which satisfies the canonical differential equations
\begin{equation}
\label{eq:CDEs}
d \mathbf{I}(\vec{x}, \epsilon) = \epsilon\, d \mathbb{A}(\vec{x})\, \mathbf{I}(\vec{x}, \epsilon)\,.
\end{equation}
The total derivative matrix $d \mathbb{A}$ can be written as a sum of $d \log$’s multiplied by constant matrices,
\begin{equation}
d \mathbb{A}(\vec{x}) = \sum_{i=1}^{n}\, \mathbb{M}_i \, d \log \eta_{i}(\vec{x})\,,
\end{equation}
where the symbol letters $\eta_i~ (i = 1, \ldots, n)$ are algebraic functions of $\vec{x}$, collectively forming the alphabet of the canonical differential system. Furthermore, the matrix $\mathbb{A}$ satisfies the integrability conditions necessary for canonical differential systems,
\begin{equation}
\partial_i \partial_j \mathbb{A} - \partial_j \partial_i \mathbb{A} = 0\,,
\qquad\quad
\big[ \partial_i \mathbb{A}\,,\, \partial_j \mathbb{A} \big]=0\,.
\end{equation}

\par
The general solution to the canonical differential equations \eqref{eq:CDEs} can be expressed in terms of Chen’s iterated integrals \cite{chen1977iterated},
\begin{equation}
\label{eq:ChenInt}
\mathbf{I}(\vec{x}, \epsilon)
= 
\mathcal{P} \exp \Big( \epsilon \int_{\gamma} d \mathbb{A} \Big)\, \mathbf{I}(\vec{x}_0, \epsilon)\,,
\end{equation}
where $\mathcal{P}$ denotes the path ordering along the integration path $\gamma$ from $\vec{x}_0$ to $\vec{x}$, and $\mathbf{I}(\vec{x}_0, \epsilon)$ is the boundary value at $\vec{x}_0$. The path-ordered exponential provides a concise representation of the following series:
\begin{equation}
\label{eq:Pexp}
\mathcal{P} \exp \Big( \epsilon \int_{\gamma} d \mathbb{A} \Big)
=
\mathbbm{1} 
+
\epsilon \int_{\gamma} d \mathbb{A}
+
\epsilon^2 \int_{\gamma} d \mathbb{A}\, d \mathbb{A}
+
\epsilon^3 \int_{\gamma} d \mathbb{A}\, d \mathbb{A}\, d \mathbb{A}
+
\cdots\,,
\end{equation}
where the $n$-th term in this expansion represents an $n$-fold iterated integral,
\begin{equation}
\int_{\gamma} \underbrace{\vphantom{,}d \mathbb{A}\, \ldots\, d \mathbb{A}}_{n \text{ times}}
=
\int_{0 \leqslant t_n \leqslant \cdots \leqslant t_1 \leqslant 1}
\mathbb{K}(t_1)\, d t_1\, \ldots\, \mathbb{K}(t_n)\, d t_n\,,
\end{equation}
with $\mathbb{K}(t)\, dt$ being the pullback of the matrix-valued differential $1$-form $d \mathbb{A}$ to the unit interval $[0, 1]$. The integrability conditions ensure that the iterated integrals in eq. \eqref{eq:Pexp} are homotopically invariant, signifying that their values are independent of the integration path, provided it avoids singularities and branch cuts of $d \mathbb{A}$. This path independence is advantageous for the analytic continuation of Feynman integrals. However, opting for a non-homotopic path can yield a different result, indicating that the Feynman integrals are multi-valued functions, yet they exhibit holomorphic behavior within each individual branch.

\par
When all symbol letters are rational functions of $\vec{x}$, the pullback of $d \mathbb{A}$ along the integration path $\gamma$ becomes a rational function of $t$ and can be further decomposed into a sum of partial fractions. Due to the $d\log$ form, each fraction features a linear denominator with respect to $t$, and is raised to a maximum power of one. Thus, by definition, all the iterated integrals in eq. \eqref{eq:Pexp}, and consequently the MIs, can be expressed using Goncharov polylogarithms (GPLs) when all symbol letters are rational functions. The GPLs are defined recursively by \cite{Goncharov:1998kja}
\begin{equation}                                                        
G(a_1, \ldots, a_n; z)
=
\int_0^z \frac{d z^{\prime}}{z^{\prime} - a_1}\, G(a_{2}, \ldots, a_n; z^{\prime})
\end{equation}
with $G(\,;z)=1$ and
\begin{equation}
G(\underbrace{0, \ldots, 0}_{n-\text{times}}; z)
=
\frac{\log^n(z)}{n!}\,,
\end{equation}
where $(a_1,\ldots, a_n)$ is referred to as the weight vector of the weight-$n$ GPL $G(a_1, \ldots, a_n; z)$. However, if the symbol letters contain square roots that cannot be rationalized simultaneously, then the MIs necessitate representation using intricate functions beyond GPLs.

\subsection{Canonical basis}
\label{sec:3.2}
The three top-level topologies illustrated in figure \ref{fig2} correspond to the following three sectors of the integral family $\mathcal{F}$:
\begin{equation}
[1, 1, 1, 0, 1, 1, 1]\,,
\qquad
[1, 0, 1, 1, 1, 1, 1]\,,
\qquad
[1, 1, 1, 1, 1, 0, 1]\,.
\end{equation}
In this subsection, we detail the construction of the canonical basis for the integral set $\mathcal{S}$ induced by the three top-sectors,
\begin{equation}
\mathcal{S}
=
[1, 1, 1, 0, 1, 1, 1]_{\digamma}
\cup\,
[1, 0, 1, 1, 1, 1, 1]_{\digamma}
\cup\,
[1, 1, 1, 1, 1, 0, 1]_{\digamma}\,,
\end{equation}
where the subscript $\digamma$ signifies the union of all the corresponding sub-sectors, defined as
\begin{equation}
[s_1, s_2, \ldots]_{\digamma}
=
\bigcup_{s_i^{\prime} \leqslant s_i}
[s_1^{\prime}, s_2^{\prime}, \ldots]\,.
\end{equation}
Following the application of IBP reduction techniques, we successfully derive a set of $32$ MIs for the integral set $\mathcal{S}$, and subsequently establish a system of linear differential equations for these MIs.

\par
\begin{figure}[htbp]
\centering
\includegraphics[width = 1.0\textwidth]{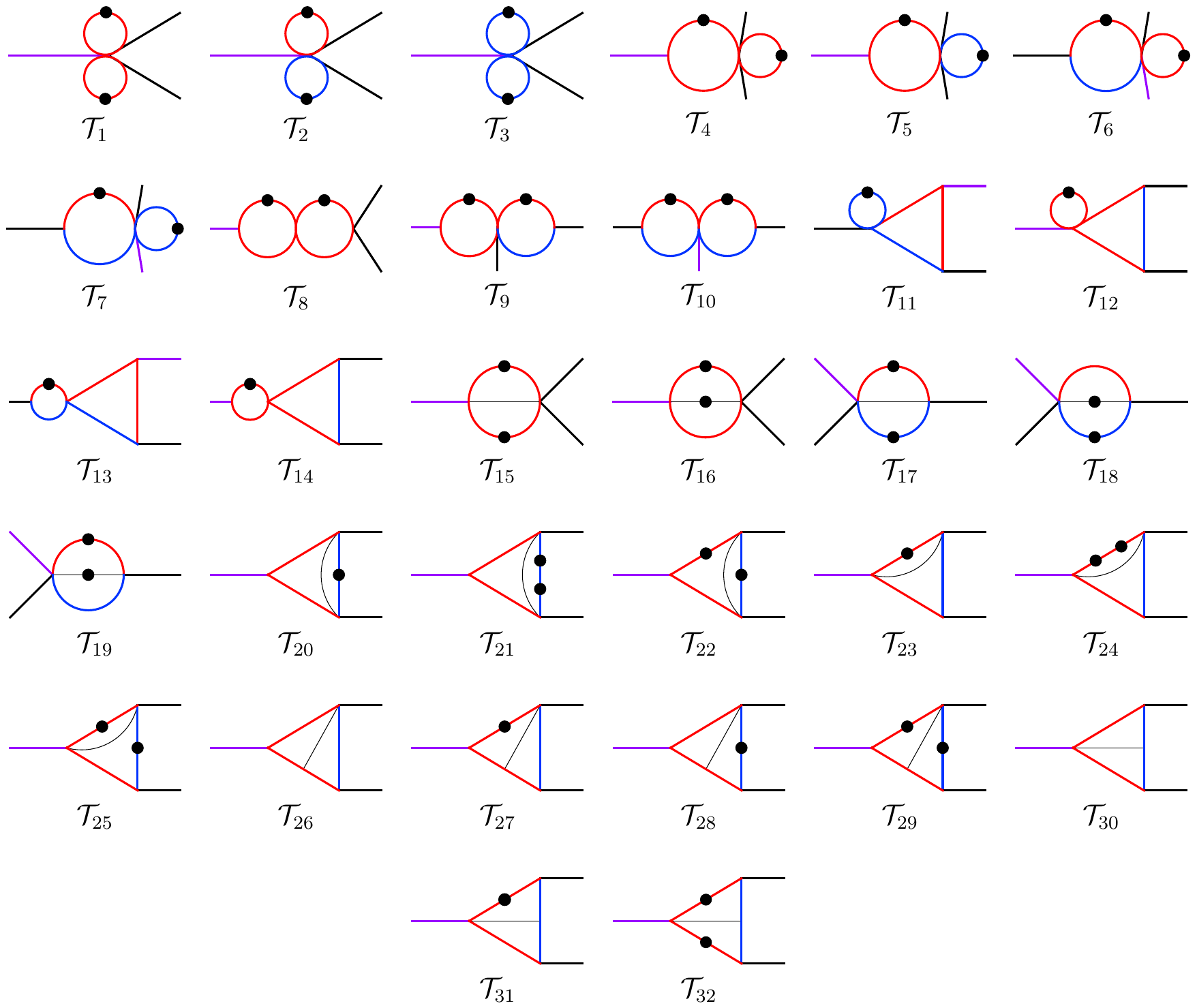} 
\caption{Pre-canonical basis of MIs for $\mathcal{S}$. The dots denote additional powers of the corresponding propagators.}
\label{fig3}
\end{figure}

We initiate the procedure with the following set of MIs,
\begin{align}
\text{F}_{1} & = \epsilon^2\, \mathcal{T}_{1}\,,  &
\text{F}_{2} & = \epsilon^2\, \mathcal{T}_{2}\,,  &
\text{F}_{3} & = \epsilon^2\, \mathcal{T}_{3}\,,  &
\text{F}_{4} & = \epsilon^2\, \mathcal{T}_{4}\,,  &
\nonumber \\
\text{F}_{5} & = \epsilon^2\, \mathcal{T}_{5}\,,  &
\text{F}_{6} & = \epsilon^2\, \mathcal{T}_{6}\,,  & 
\text{F}_{7} & = \epsilon^2\, \mathcal{T}_{7}\,,  &
\text{F}_{8} & = \epsilon^2\, \mathcal{T}_{8}\,,  &
\nonumber \\
\text{F}_{9} & = \epsilon^2\, \mathcal{T}_{9}\,,  & 
\text{F}_{10} & = \epsilon^2\, \mathcal{T}_{10}\,,  &
\text{F}_{11} & = \epsilon^3\, \mathcal{T}_{11}\,,  &
\text{F}_{12} & = \epsilon^3\, \mathcal{T}_{12}\,,  &
\nonumber \\
\text{F}_{13} & = \epsilon^3\, \mathcal{T}_{13}\,,  &
\text{F}_{14} & = \epsilon^3\, \mathcal{T}_{14}\,,  &
\text{F}_{15} & = \epsilon^2\, \mathcal{T}_{15}\,,  & 
\text{F}_{16} & = \epsilon^2\, \mathcal{T}_{16}\,,  &
\nonumber \\
\text{F}_{17} & = \epsilon^2\, \mathcal{T}_{17}\,,  &
\text{F}_{18} & = \epsilon^2\, \mathcal{T}_{18}\,,  & 
\text{F}_{19} & = \epsilon^2\, \mathcal{T}_{19}\,,  &
\text{F}_{20} & = \epsilon^3\, \mathcal{T}_{20}\,,
\\
\text{F}_{21} & = \epsilon^2\, \mathcal{T}_{21}\,,  & 
\text{F}_{22} & = \epsilon^2\, \mathcal{T}_{22}\,,  &
\text{F}_{23} & = \epsilon^3\, \mathcal{T}_{23}\,,  &
\text{F}_{24} & = \epsilon^2\, \mathcal{T}_{24}\,,  &
\nonumber \\
\text{F}_{25} & = \epsilon^2\, \mathcal{T}_{25}\,,  &
\text{F}_{26} & = \epsilon^4\, \mathcal{T}_{26}\,,  &
\text{F}_{27} & = \epsilon^3\, \mathcal{T}_{27}\,,  & 
\text{F}_{28} & = \epsilon^3\, \mathcal{T}_{28}\,,  &
\nonumber \\
\text{F}_{29} & = \epsilon^2\, \mathcal{T}_{29}\,,  &
\text{F}_{30} & = \epsilon^4\, \mathcal{T}_{30}\,,  & 
\text{F}_{31} & = \epsilon^3\, \mathcal{T}_{31}\,,  &
\text{F}_{32} & = \epsilon^2\, \mathcal{T}_{32}\,,  &
\nonumber
\end{align}
where $\mathcal{T}_i~ (i = 1, \ldots, 32)$ are illustrated in figure \ref{fig3}. This set of MIs satisfies a linear-form differential system,
\begin{equation}
d \mathbf{F}(\vec{x}, \epsilon)
=
d \Big[ \mathbb{A}^{(0)}(\vec{x}) + \epsilon\, \mathbb{A}^{(1)}(\vec{x}) \Big]\,
\mathbf{F}(\vec{x}, \epsilon)\,,
\end{equation}
and is conventionally referred to as a linear basis of MIs. Following the algorithm suggested in ref. \cite{DiVita:2014pza}, we employ the Magnus exponential method to construct a set of canonical MIs, adhering to the canonical differential equations \eqref{eq:CDEs},
{\setlength{\lineskip}{3pt}
\setlength{\lineskiplimit}{4pt}
\begin{align}
\label{eq:UTbasis}
\text{I}_{1} & = \text{F}_{1}\,,  &
\text{I}_{2} & = \text{F}_{2}\,,
\nonumber \\
\text{I}_{3} & = \text{F}_{3}\,,  &
\text{I}_{4} & = \lambda_1\, \text{F}_{4}\,,
\nonumber \\
\text{I}_{5} & = \lambda_1\, \text{F}_{5}\,,
\nonumber \\
\text{I}_{6} & = \frac{\lambda_3}{2\, (m_t^2 - m_b^2 - m_{\scriptscriptstyle{W}}^2)}\,
                        (\text{F}_{1} - \text{F}_{2} - 2\, m_{\scriptscriptstyle{W}}^2\, \text{F}_{6})\,,
\nonumber \\
\text{I}_{7} & = \frac{\lambda_3}{2\, (m_t^2 - m_b^2 - m_{\scriptscriptstyle{W}}^2)}\,
                        (\text{F}_{2} - \text{F}_{3} - 2\, m_{\scriptscriptstyle{W}}^2\, \text{F}_{7})\,,  &
\text{I}_{8} & = \lambda_1^2\, \text{F}_{8}\,,
\nonumber \\
\text{I}_{9} & = \frac{\lambda_1\, \lambda_3}{2\, (m_t^2 - m_b^2 - m_{\scriptscriptstyle{W}}^2)}\,
                        (\text{F}_{4} - \text{F}_{5} - 2\, m_{\scriptscriptstyle{W}}^2\, \text{F}_{9})\,,
\nonumber \\ 
\text{I}_{10} & = \rlap{$\displaystyle
                          \frac{\lambda_3^2}{4\, (m_t^2 - m_b^2 - m_{\scriptscriptstyle{W}}^2)^2}\,
                          ( \text{F}_{1} - 2\, \text{F}_{2} + \text{F}_{3} - 4\, m_{\scriptscriptstyle{W}}^2\, \text{F}_{6} + 4\, m_{\scriptscriptstyle{W}}^2\, \text{F}_{7} + 4\, m_{\scriptscriptstyle{W}}^4\, \text{F}_{10} )\,,$}
\nonumber \\
\text{I}_{11} & = \lambda_2\, \text{F}_{11}\,,  &
\text{I}_{12} & = \lambda_2\, \text{F}_{12}\,,
\nonumber \\
\text{I}_{13} & = \frac{\lambda_2\, \lambda_3}{2\, (m_t^2 - m_b^2 - m_{\scriptscriptstyle{W}}^2)}\,
                          (-\, \text{F}_{11} + \text{F}_{12} - 2\, m_{\scriptscriptstyle{W}}^2\, \text{F}_{13})\,,  &
\text{I}_{14} & = \lambda_1\, \lambda_2\, \text{F}_{14}\,,
\nonumber \\
\text{I}_{15} & = s\, \text{F}_{15}\,,  &
\text{I}_{16} & = \frac{\lambda_1}{2}\, (\text{F}_{15} + 2\, \text{F}_{16})\,,
\nonumber \\
\text{I}_{17} & = m_{\scriptscriptstyle{W}}^2\, \text{F}_{17} \,,  &
\text{I}_{18} & = \lambda_3\, (\text{F}_{17} + \text{F}_{18} + \text{F}_{19})\,,
\nonumber \\
\text{I}_{19} & = \rlap{$\displaystyle
                          \frac{1}{4}\, (2\, m_t^2 - 2\, m_b^2 - m_{\scriptscriptstyle{W}}^2)\, \text{F}_{17}
                       + \frac{1}{2}\, (m_t^2 - m_b^2 - m_{\scriptscriptstyle{W}}^2)\, \text{F}_{18}
                       + \frac{1}{2}\, (m_t^2 - m_b^2 + m_{\scriptscriptstyle{W}}^2)\, \text{F}_{19}\,,$}
\nonumber \\
\text{I}_{20} & = \lambda_2\, \text{F}_{20}\,,  &
\text{I}_{21} & = \lambda_2\, m_b^2\, \text{F}_{21}\,,
\nonumber \\
\text{I}_{22} & = \lambda_1\, \big[\,
                          \frac{3}{2}\, \text{F}_{20} + m_b^2\, \text{F}_{21} + (m_t^2 - m_{\scriptscriptstyle{W}}^2)\, \text{F}_{22}\,
                          \big]\,,  &
\text{I}_{23} & = \lambda_2\, \text{F}_{23}\,,
\nonumber \\
\text{I}_{24} & = \lambda_2\, m_t^2\, \text{F}_{24}\,,
\\
\text{I}_{25} & = \rlap{$\displaystyle
                          \frac{\lambda_3}
                          {s\, (m_t^2 + m_b^2 - m_{\scriptscriptstyle{W}}^2) - m_t^2\, (m_t^2 - m_b^2 + m_{\scriptscriptstyle{W}}^2)}\,
                          \Big\{
                          \frac{1}{4}\, (s - m_t^2)\, (-\, \text{F}_{1} + \text{F}_{2} + 3\, s\, \text{F}_{15})$}
                          \nonumber \\
                    &    \quad~
                          \rlap{$\displaystyle
                          -\, \frac{1}{8}\,
                          \big[\,
                          s\, (m_t^2 - 3\, m_b^2 + 7\, m_{\scriptscriptstyle{W}}^2)
                          - m_t^2\, (m_t^2 - m_b^2 + 5\, m_{\scriptscriptstyle{W}}^2)
                          \,\big]\,
                          \text{F}_{17}$}
                          \nonumber \\
                    &    \quad~
                          \rlap{$\displaystyle
                          -\, \frac{1}{8}\,
                          \big[\,
                          s\, (m_t^2 - 3\, m_b^2 - m_{\scriptscriptstyle{W}}^2)
                          - m_t^2\, (m_t^2 - m_b^2 - 3\, m_{\scriptscriptstyle{W}}^2)
                          \,\big]\,
                          \text{F}_{18}$}
                          \nonumber \\
                    &   \quad~
                          \rlap{$\displaystyle
                          -\, \frac{1}{8}\,
                          \big[\,
                          s\, (m_t^2 - 3\, m_b^2 + 3\, m_{\scriptscriptstyle{W}}^2)
                          - m_t^2\, (m_t^2 - m_b^2 + m_{\scriptscriptstyle{W}}^2)
                          \,\big]\,
                          \text{F}_{19} $}
                          \nonumber \\
                    &    \quad~
                          \rlap{$\displaystyle
                          +\, \frac{3}{2}\, s\, (s - m_t^2 + 2\, m_b^2 - 2\, m_{\scriptscriptstyle{W}}^2)\, \text{F}_{23}
                          + s\, m_t^2\, (s - m_t^2 + 2\, m_b^2 - 2\, m_{\scriptscriptstyle{W}}^2)\, \text{F}_{24}$}
                          \nonumber \\
                    &    \quad~
                          \rlap{$\displaystyle
                          +\, \big[\,
                          s^2\, m_b^2
                          - s\, m_t^2\, (m_b^2 + m_{\scriptscriptstyle{W}}^2)
                          + s\, (m_b^2 - m_{\scriptscriptstyle{W}}^2)^2
                          + m_t^4\, m_{\scriptscriptstyle{W}}^2 \,\big]\,
                          \text{F}_{25}
                          \Big\}\,,$}
\nonumber \\
\text{I}_{26} & = \lambda_2\, \text{F}_{26}\,,  &
\text{I}_{27} & = \lambda_1\, \lambda_2\, \text{F}_{27}\,,
\nonumber \\
\text{I}_{28} & = \lambda_2\, \lambda_3\, \text{F}_{28}\,,
\nonumber \\
\text{I}_{29} & = \rlap{$\displaystyle
                          s\, \frac{m_t^2 + m_b^2 - m_{\scriptscriptstyle{W}}^2}
                          {m_t^2 - m_b^2 - m_{\scriptscriptstyle{W}}^2}\,
                          (\text{F}_{4} - \text{F}_{5} - 2\, m_{\scriptscriptstyle{W}}^2\, \text{F}_{9})
                          + s\, (s - 2\, m_t^2 + 2\, m_b^2 - 2\, m_{\scriptscriptstyle{W}}^2)\, \text{F}_{27}$}
                          \nonumber \\
                    &   \quad~
                         \rlap{$\displaystyle
                         +\, s\, (m_t^2 - m_b^2 - m_{\scriptscriptstyle{W}}^2)\, \text{F}_{28}
                         + s\, \big[\,
                         (m_t^2 - m_{\scriptscriptstyle{W}}^2)^2
                         + m_b^2\, (s - 2\, m_t^2 - 2\, m_{\scriptscriptstyle{W}}^2)
                         + m_b^4
                         \,\big]\, \text{F}_{29}\,,$}
\nonumber \\
\text{I}_{30} & = \lambda_2\, \text{F}_{30}\,,  &
\text{I}_{31} & = \lambda_2\, \lambda_3\, \text{F}_{31}\,,
\nonumber \\
\text{I}_{32} & = \rlap{$\displaystyle
                          -\, \frac{s\, m_b^2}{(m_t^2 - m_b^2 - m_{\scriptscriptstyle{W}}^2)^2}\,
                          (\text{F}_{1} - 2\, \text{F}_{2} + \text{F}_{3}
                          - 4\, m_{\scriptscriptstyle{W}}^2\, \text{F}_{6}
                          + 4\, m_{\scriptscriptstyle{W}}^2\, \text{F}_{7}
                          + 4\, m_{\scriptscriptstyle{W}}^4\, \text{F}_{10})$}
                          \nonumber \\ 
                    &    \quad~
                          \rlap{$\displaystyle
                          +\, 2\, s\, (m_t^2 - m_b^2 - m_{\scriptscriptstyle{W}}^2)\, \text{F}_{31}
                          + s\, \big[\,
                          (m_t^2 - m_{\scriptscriptstyle{W}}^2)^2
                          + m_b^2\, (s - 2\, m_t^2 - 2\, m_{\scriptscriptstyle{W}}^2)
                          + m_b^4 \,\big]\, \text{F}_{32}\,,$}
\nonumber
\end{align}
}
where $\lambda_{1, 2, 3}$ are square roots related to kinematics, defined as
\begin{equation}
\lambda_1^2 = s\, (s - 4\, m_t^2)\,,
\qquad\quad
\lambda_2^2 = s\, (s - 4\, m_{\scriptscriptstyle{W}}^2)\,,
\qquad\quad
\lambda_3^2 = \lambda(m_{\scriptscriptstyle{W}}^2,\, m_t^2,\, m_b^2)\,,
\end{equation}
and $\lambda(x, y, z)$ therein represents the K{\"a}ll{\'e}n function,
\begin{equation}
\lambda(x, y, z) = x^2 + y^2 + z^2 - 2\, x\, y - 2\, y\, z - 2\, z\, x\,.
\end{equation}
All canonical MIs specified in eq. \eqref{eq:UTbasis} are normalized to ensure finiteness as $\epsilon \rightarrow 0$. Dimensional analysis confirms that each of these MIs is dimensionless.

\par
To facilitate our subsequent discussion, we introduce the following three dimensionless variables,
\begin{equation}
\tau_1 = -\, \frac{s}{m_t^2}\,,
\qquad\quad
\tau_2 = -\, \frac{m_{\scriptscriptstyle{W}}^2}{m_t^2}\,,
\qquad\quad
\tau_3 = \frac{m_b^2}{m_t^2}\,.
\end{equation}
The canonical differential equations of $\mathbf{I}$ with respect to $\tau_1,\tau_2$ and $\tau_3$ are dependent on the reduced square roots $\bar{\lambda}_i \equiv \lambda_i/m_t^2~ (i = 1, 2, 3)$,
\begin{equation}
\bar{\lambda}_1^2 = \tau_1\, (\tau_1 + 4) \,,
\qquad\quad
\bar{\lambda}_2^2 = \tau_1\, (\tau_1 - 4\, \tau_2) \,,
\qquad\quad
\bar{\lambda}_3^2 = \lambda(1,\, -\, \tau_2,\, \tau_3)\,. 
\end{equation}
Aided by the \texttt{RationalizeRoots} package \cite{Besier:2019kco}, we successfully rationalize $\bar{\lambda}_1$, $\bar{\lambda}_2$ and $\bar{\lambda}_3$ simultaneously as
\begin{equation}
\bar{\lambda}_1 = \frac{(1 - x)\, (1 + x)}{x}\,,
\qquad
\bar{\lambda}_2 = \frac{(1 - x)^2\, (1 - z)}{x\, (1 + z)}\,,
\qquad
\bar{\lambda}_3 = \frac{(1 - x)^2\, z + x\, y^2}{x\, y\, (1 + z)}
\end{equation}
by the following change of variables,
\begin{equation}
(\tau_1, \tau_2, \tau_3)
\longmapsto
(x, y, z):
\quad
\left\lbrace~
\begin{aligned}
&
\tau_1 = \frac{(1 - x)^2}{x}
\\
&
\tau_2 = \frac{(1 - x)^2\, z}{x\, (1 + z)^2}
\\
&
\tau_3 = \Big(1 + \frac{y}{1 + z} \Big)\, \Big[\, 1 - \frac{(1 - x)^2\, z}{x\, y\, (1 + z)} \,\Big]
\end{aligned}
\right.
\end{equation}
Consequently, the differential system of the integral set $\mathcal{S}$ is cast into a $d\log$ form with rational symbol letters,
\begin{equation}
d \mathbf{I}(x, y, z, \epsilon)
=
\epsilon\,
\Big[\, \sum_{i=1}^{20}\, \mathbb{M}_i \, d \log \eta_{i}(x, y, z) \,\Big]\,
\mathbf{I}(x, y, z, \epsilon)\,,
\end{equation}
where the symbol letters $\eta_i$ are given as
\begin{align}
\eta_{1} & = x\,,  &
\eta_{11} & = 1 + x\, z\,,
\nonumber \\
\eta_{2} & = y\,,  &
\eta_{12} & = 1 - x + y\,,
\nonumber \\
\eta_{3} & = z\,,  &
\eta_{13} & = - 1 + x + x\, y\,,
\nonumber \\
\eta_{4} & = 1 - x\,,  &
\eta_{14} & = 1 + y + z\,,
\nonumber \\
\eta_{5} & = 1 + x\,,  &
\eta_{15} & = y + (1 - x)\, z\,,
\\
\eta_{6} & = 1 + y\,,  &
\eta_{16} & = x\, y - (1 - x)\, z\,,
\nonumber \\
\eta_{7} & = 1 - z\,,  &
\eta_{17} & = x\, y - (1 - x)^2\,,
\nonumber \\
\eta_{8} & = 1 + z\,,  &
\eta_{18} & = x\, y - (1 - x)^2\, z\,,
\nonumber \\
\eta_{9} & = x + z\,,  &
\eta_{19} & = x\, y^2 + (1 - x)^2\, z\,,
\nonumber \\
\eta_{10} & = y + z\,,  &
\eta_{20} & = x\, y\, (1 + z) - (1 - x)^2\, z\,,
\nonumber
\end{align}
and the explicit expressions of the coefficient matrices $\mathbb{M}_i$ are presented in the supplementary file ``dlog-form\_Matrix.m." In the designated positive-letter region, defined by
\begin{equation}
\label{eq:PLregion}
0 < x < 1
\quad
\wedge
\quad
y > \frac{1}{x}
\quad
\wedge
\quad
0 < z < 1\,,
\end{equation}
all symbol letters are real and positive.

\par
The Euclidean region for this system is delineated by the following kinematic constraints:
\begin{equation}
\label{eq:Eregion}
s < 0\,,
\qquad\quad
m_{\scriptscriptstyle{W}}^2 < 0\,,
\qquad\quad
m_t^2 > 0\,,
\qquad\quad
m_b^2 > 0\,.
\end{equation}
In the positive-Euclidean region, defined as the intersection of the positive-letter and Euclidean regions, the canonical MIs $\text{I}_i$ are real functions of $x$, $y$ and $z$, and can be concisely expressed in terms of GPLs. To perform calculations for the mixed QCD-EW corrections to the $e^+e^- \rightarrow W^+W^-$ process, it is necessary to analytically continue these MIs to the physical region.

\subsection{Boundary conditions}
\label{sec:3.3}
\par
To arrive at a definite solution for the canonical differential system \eqref{eq:CDEs}, it is essential to specify boundary conditions. At present, there is no comprehensive algorithm or tool that can automate the process of establishing boundary conditions. However, two strategies are commonly employed to determine boundary constants:
\begin{itemize}
\item Strategy 1: Fixing the boundary constants by an independent, and often simpler, calculation at a certain preferred kinematic point. This specific point may possess some accidental kinematic symmetries or have particular physical significance. Typically, the number of independent MIs needed at this point is reduced dramatically due to the degeneracy of the MIs at this kinematic limit. Various analytical and numerical techniques can then be applied to evaluate these MIs directly at the chosen point, including Feynman parametrization with Cheng-Wu theorem \cite{Cheng:1987ga}, Mellin-Barnes representation \cite{Smirnov:1999gc,Tausk:1999vh}, expansion by regions \cite{Smirnov:1994tg,Beneke:1997zp,Jantzen:2011nz}, sector decomposition \cite{Heinrich:2008si,Borowka:2017idc,Smirnov:2021rhf}, and the auxiliary mass flow method \cite{Liu:2017jxz,Liu:2022mfb,Liu:2022chg}.
\item Strategy 2: Ascertaining the boundary constants by formulating an Ansatz based on the asymptotic behavior of the MIs around the singularities of the differential equations. The typical approach is to stipulate the regularity of the MIs or their linear combinations at spurious singularities, from which a set of linear equations can be systematically constructed for the undetermined boundary constants.
\end{itemize}

\par
\begin{figure}[htbp]
\centering
\includegraphics[width = 1.0\textwidth]{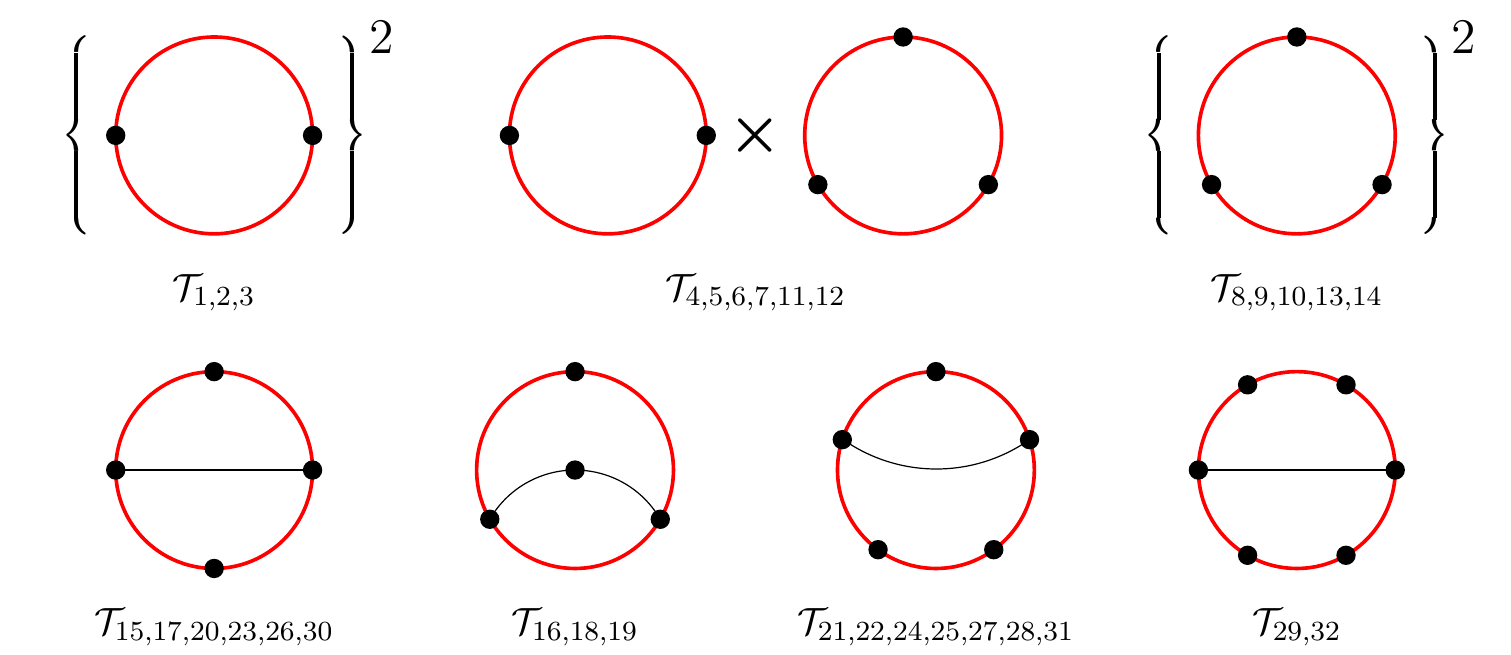} 
\caption{Seven vacuum integrals induced by pre-canonical MIs $\mathcal{T}_{i}~ (i = 1, \ldots, 32)$ at $\vec{x}_0 = (1, 0, 0)$.
}
\label{fig4}
\end{figure}

In addressing our specific problem, we determine the boundary constants by adopting the first of the aforementioned strategies, evaluating the canonical MIs at the kinematic point $\vec{x}_0 = (1, 0, 0)$. At this point, the kinematic variables simplify to
\begin{equation}
s = 0\,,
\qquad\quad
m_{\scriptscriptstyle{W}}^2 = 0\,,
\qquad\quad
m_b^2 = m_t^2\,.
\end{equation}
For this kinematic configuration, all pre-canonical MIs $\mathcal{T}_{i}$ degenerate into the following seven vacuum integrals,
\begin{align}
& \mathcal{V}_{1} = \mathcal{T}_{1, 2, 3}\,, &
& \mathcal{V}_{2} = \mathcal{T}_{4, 5, 6, 7, 11, 12}\,, &
& \mathcal{V}_{3} = \mathcal{T}_{8, 9, 10, 13, 14}\,,
\nonumber \\
& \mathcal{V}_{4} = \mathcal{T}_{15, 17, 20, 23, 26, 30}\,, &
& \mathcal{V}_{5} = \mathcal{T}_{16, 18, 19}\,, &
& \mathcal{V}_{6} = \mathcal{T}_{21, 22, 24, 25, 27, 28, 31}\,,
\\
& \mathcal{V}_{7} = \mathcal{T}_{29, 32}\,,
\nonumber
\end{align}
which are graphically illustrated in figure \ref{fig4}. Additionally, being part of the same integral family, these vacuum integrals can be reduced to a single independent MI by employing IBP recurrence relations,
\begin{align}
\frac{\mathcal{V}_{2}}{\mathcal{V}_{1}} & = -\, \frac{\epsilon}{2\, m_t^2}\,, &
\frac{\mathcal{V}_{3}}{\mathcal{V}_{1}} & = \frac{\epsilon^2}{4\, m_t^4}\,,
\nonumber \\
\frac{\mathcal{V}_{4}}{\mathcal{V}_{1}} & = -\, \frac{\epsilon^2}{(1 - \epsilon)\, (1 + 2\, \epsilon)\, m_t^2}\,, &
\frac{\mathcal{V}_{5}}{\mathcal{V}_{1}} & = \frac{\epsilon}{2\, (1 - \epsilon)\, (1 + 2\, \epsilon)\, m_t^2}\,,
\\
\frac{\mathcal{V}_{6}}{\mathcal{V}_{1}} & = \frac{\epsilon^2}{4\, (1 - \epsilon)\, m_t^4}\,, &
\frac{\mathcal{V}_{7}}{\mathcal{V}_{1}} & = -\, \frac{\epsilon^2\, (1 + \epsilon)^2}{4\, (1 - \epsilon)\, (3 + 2\, \epsilon)\, m_t^6}\,.
\nonumber
\end{align}
All canonical MIs $\text{I}_i~ (i = 1, \ldots, 32)$ are regular at $\vec{x}_0$. A straightforward analysis of the transformation \eqref{eq:UTbasis} allows us to deduce that
\begin{equation}
\text{I}_i(\vec{x}_0, \epsilon)
=
\big( \delta_{i1} + \delta_{i2} + \delta_{i3} \big)\, \epsilon^2\, \mathcal{V}_1\,,
\end{equation}
where $\mathcal{V}_1 = 1/\epsilon^2$, computed directly using Feynman parametrization.

\par
\begin{figure}[htbp]
\centering
\includegraphics[width = 0.45\textwidth]{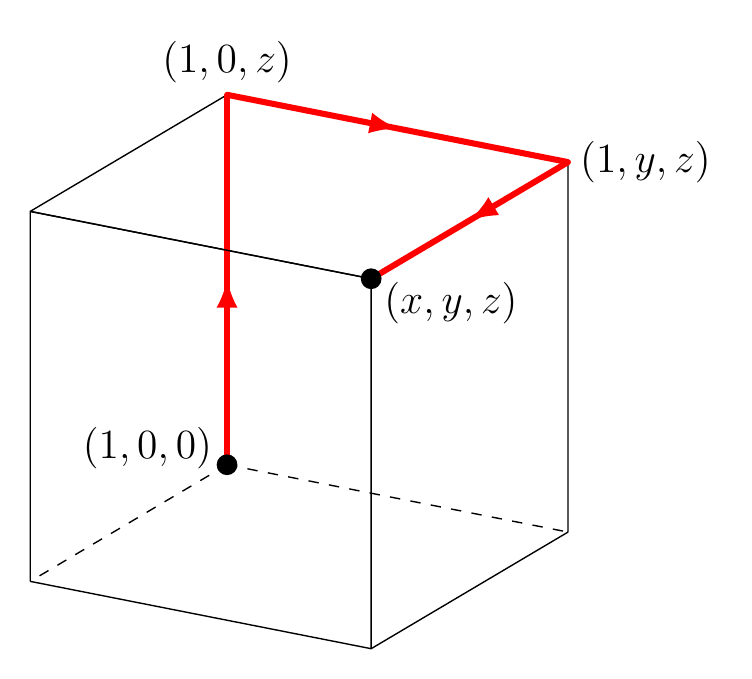} 
\caption{Integration path connecting $\vec{x}_0 = (1, 0, 0)$ to $\vec{x} = (x, y, z)$.}
\label{fig5}
\end{figure}

Upon performing the path-ordered integration \eqref{eq:ChenInt}, we derive the analytic expressions for all canonical MIs $\text{I}_i$, formulated in terms of GPLs. The chosen integration path $\gamma$, which connects the start point $\vec{x}_0 = (1, 0, 0)$ to the point of interest $\vec{x} = (x, y, z)$, is a piecewise linear path that proceeds from $\vec{x}_0$ to $(1, 0, z)$, then to $(1, y, z)$, and finally to $\vec{x}$, as illustrated in figure \ref{fig5}. Consequently, the arguments of the GPLs involved in the canonical MIs are $x$, $y$ or $z$, and the weights of these GPLs are the zeros of the symbol letters, cataloged in table \ref{table:weights}. For the symbolic computation and numerical evaluation of GPLs, we utilize the \texttt{Mathematica} package \texttt{PolyLogTools} \cite{Maitre:2005uu,Maitre:2007kp,Duhr:2019tlz} and \texttt{C++}  library \texttt{GiNaC} \cite{Bauer:2000cp,Vollinga:2004sn}. To ensure the accuracy and reliability of our analytic expressions, we perform numerical cross-checks with extremely high precision for all MIs within the positive-Euclidean region. This validation is conducted using the publicly available packages \texttt{pySecDec} and \texttt{AMFlow}. In the appendix,  we showcase the explicit expressions of $\text{I}_{1, \ldots, 32}$ up to $\mathcal{O}(\epsilon^2)$. The analytic expressions of these canonical MIs up to $\mathcal{O}(\epsilon^4)$, which are essential for the NNLO mixed QCD-EW corrections to $e^+e^- \rightarrow W^+W^-$, are available in the supplementary file ``analytic\_MIs.m," accompanying the arXiv submission of this paper.
\begin{table}[htbp]
\centering
\renewcommand{\arraystretch}{1.4}
\begin{tabular}{
|
p{2.1cm}<{\centering}
|
p{12cm}
|
}
\toprule[1.2pt]
\multirow{1}{*}{GPL}
&
\qquad\qquad\qquad\qquad\qquad\qquad\quad
\multirow{1}{*}{Weight}
\\
\midrule[0.4pt]
\multirow{1}{*}{$G(a_1, \ldots ; x)$}
&\,
\multirow{1}{*}{$-\,1\,, \quad 0\,, \quad 1\,,$}
\\
\multirow{1}{*}{$G(a_1, \ldots ; y)$}
&\,
\multirow{1}{*}{$-\,1\,, \quad 0\,, \quad -\,(1-x)\,, \quad (1-x)/x\,, \quad (1-x)^2/x\,,$}
\\
\multirow{2}{*}{$G(a_1, \ldots ; z)$}
&\,
\multirow{1}{*}{$-\,1\,, \quad 0\,, \quad 1\,, \quad -\,x\,, \quad -\,y\,, \quad -\,(1+y)\,, \quad -\,1/x\,, \quad -\,y/(1-x)\,,$}
\\
&\,
\multirow{1}{*}{$x\,y/(1-x)\,, \quad x\,y/(1-x)^2\,, \quad -\,x\,y^2/(1-x)^2\,, \quad x\,y/[\,(1-x)^2-x\,y\,]$}
\\
\bottomrule[1.2pt]
\end{tabular}
\caption{Weights of GPLs involved in $\text{I}_i~ (i = 1, \ldots, 32)$.}
\label{table:weights}
\end{table}

\subsection{Analytic continuation}
\label{sec:3.4}
\par
We have successfully derived the analytic expressions for the canonical MIs within the positive-Euclidean region specified by eqs. \eqref{eq:PLregion} and \eqref{eq:Eregion}. The task that now remains is to analytically continue these MIs from the positive-Euclidean region to the physical region \cite{Gehrmann:2013cxs,Bonciani:2016ypc,DiVita:2017xlr}. Consequently, it is necessary to carry out the analytic continuation for our chosen kinematic variables $x$, $y$ and $z$. This entails applying the Feynman prescription to both external and internal Lorentz invariants. Specifically, in the positive-Euclidean region, the dimensionless variables $x$, $y$ and $z$ can be formulated as
\begin{equation}
\label{eq:xyz}
x = \frac{\sqrt{4\, m_t^2 - s} - \sqrt{-\, s}}{\sqrt{4\, m_t^2 - s} + \sqrt{-\, s}}\,,
\qquad\quad
z = \frac{\sqrt{-\, s} - \sqrt{4\, m_{\scriptscriptstyle{W}}^2 - s}}{\sqrt{-\, s} + \sqrt{4\, m_{\scriptscriptstyle{W}}^2 - s}}\,,
\qquad\quad
y = (1 + z)\, \xi
\end{equation}
with
\begin{equation}
\xi = \frac{1}{2\, m_t^2}\, \Big[\, m_b^2 - m_t^2 - m_{\scriptscriptstyle{W}}^2 + \sqrt{\lambda(m_{\scriptscriptstyle{W}}^2,\, m_t^2,\, m_b^2)} \,\Big]\,.
\end{equation}
Following the Feynman prescription, we assign a positive infinitesimal imaginary part to the external Mandelstam invariants and a negative infinitesimal imaginary part to the internal mass squares in the following manner:
\begin{equation}
\label{eq:F-prescription}
s + i\, 0^+\,,
\qquad\quad
m_{\scriptscriptstyle{W}}^2 + i\, 0^+\,,
\qquad\quad
m_b^2 - i\, 0^+\,,
\qquad\quad
m_t^2 - i\, 0^+\,.
\end{equation}

\par
To begin with, we focus on the analytic continuation of the variable $x$. In this procedure, it is crucial to select the correct branch for each square root, in accordance with the Feynman prescription given in eq. \eqref{eq:F-prescription}. The continuation of $x$ from the positive-Euclidean region to the full domain of $\tau_1$ is established as follows:\footnote{Due to the integration measure specified in eq. \eqref{eq:measure}, $m_t^2$ is restricted to be positive throughout the analytic continuation procedure.}
\begin{equation}
x =
\left\lbrace~
\begin{aligned}
&
\frac{1}{2}\, \big[\, \tau_1 + 2 - \sqrt{\tau_1\, (\tau_1 + 4)} \,\big]
&\qquad&
\tau_1 \in (0,\, +\, \infty)
\\
&\,
\vphantom{\frac{1}{1}}{e^{\, i \vartheta}}
&\qquad&
\tau_1 \in (-\, 4,\, 0)
\\
&
\frac{1}{2}\, \big[\, \tau_1 + 2 + \sqrt{\tau_1\, (\tau_1 + 4)} \,\big] + i\, 0^+
&\qquad&
\tau_1 \in (-\, \infty,\,-\, 4)
\end{aligned}
\right.
\end{equation}
where $\vartheta = 2 \arctan\sqrt{-\, \tau_1/(\tau_1 + 4)}$, and the three distinct regions of $\tau_1$ are identified based on the signs of the real and imaginary parts of $x$. In the positive-Euclidean region where $\tau_1 > 0$, $x$ is real and varies within $(0,\, 1)$. When $-\, 4 < \tau_1 < 0$, $x$ assumes a pure phase. In the case where $\tau_1 < -\, 4$, $x$ is negative, accompanied by a positive infinitesimal imaginary part.

\par
Similar to the treatment of $x$, the variable $z$ is continued to the entire $(r,\, \tau_1)$-plane,
\begin{equation}
\label{eq:z-continuation}
z =
\left\lbrace~
\begin{aligned}
&
\frac{1}{2}\, \big[\, r - 2 - \sqrt{r\, (r - 4)} \,\big]
&\qquad&
\text{I}:
&&
r \in (4,\, +\, \infty)
&&
\\
&\,
\vphantom{\frac{1}{1}}{e^{\, - i \varphi}}
&\qquad&
\text{II}:
&&
r \in (0,\, 4)
&&
\\
&
\frac{1}{2}\, \big[\, r - 2 + \sqrt{r\, (r - 4)} \,\big] - i\, 0^+
&\qquad&
\text{IIIa}:
&&
r \in (-\, \infty,\, 0)\,,
&&
\tau_1 > 0
\\
&
\frac{1}{2}\, \big[\, r - 2 - \sqrt{r\, (r - 4)} \,\big] - i\, 0^+
&\qquad&
\text{IIIb}:
&&
r \in (-\, 4,\, 0)\,,
&&
\tau_1 < 0
\\
&
\frac{1}{2}\, \big[\, r - 2 + \sqrt{r\, (r - 4)} \,\big] + i\, 0^+
&\qquad&
\text{IIIc}:
&&
r \in (-\, \infty,\, -\, 4)\,,
&&
\tau_1 < 0
\end{aligned}
\right.
\end{equation}
where $r = \tau_1/\tau_2 = s/m_{\scriptscriptstyle{W}}^2$ and $\varphi = 2 \arctan\sqrt{-\, (r - 4)/r}$. Based on these results, the $(r,\, \tau_1)$-plane is partitioned into five distinct regions. In region I, $z$ is real and ranges from $0$ to $1$. In region II, $z$ becomes a pure phase. In the remaining three regions, $z$ is negative with an infinitesimal imaginary part, the sign of which is specified in eq. \eqref{eq:z-continuation}.

\par
\begin{figure}[htbp]
\centering
\includegraphics[width = 0.55\textwidth]{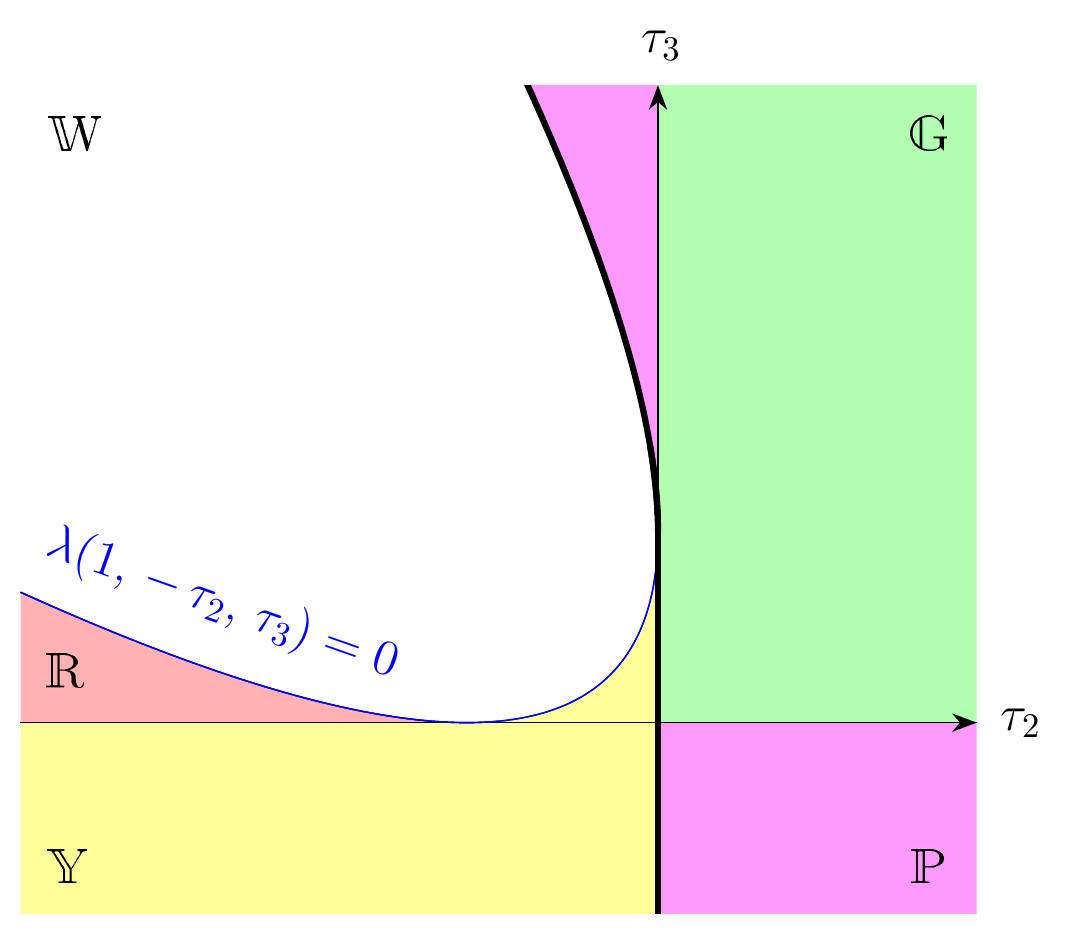} 
\caption{
Five distinct regions for the analytic continuation of $\xi$. In the white region ($\mathbbm{W}$), $\xi$ has a finite imaginary part, whereas in the multi-colored area, the imaginary part of $\xi$ is either zero or infinitesimal. The bold line represents the boundary where the real part of $\xi$ changes sign. To the left of this demarcation, in the red ($\mathbbm{R}$) and yellow ($\mathbbm{Y}$) regions, $\mathrm{Re}\, \xi < 0$; to the right, in the green ($\mathbbm{G}$) and purple ($\mathbbm{P}$) regions, $\mathrm{Re}\, \xi > 0$.
}
\label{fig6}
\end{figure}

As indicated in eq. \eqref{eq:xyz}, to perform the analytic continuation of $y$, we decompose it into the factors $(1 + z)$ and $\xi$. The continuation of the first factor can be inferred from eq. \eqref{eq:z-continuation}; our focus here is on the continuation of the latter factor. Following the Feynman prescription for $m_{\scriptscriptstyle{W}}^2$, $m_t^2$ and $m_b^2$, the analytic continuation of $\xi$ can be precisely delineated as
\begin{equation}
\xi =
\left\lbrace~
\begin{aligned}
&
\frac{1}{2}\, \big[\, \tau_2 + \tau_3 - 1 +  \sqrt{\lambda(1,\, -\, \tau_2,\, \tau_3)} \,\big]
&\qquad&
(\tau_2,\, \tau_3) \in \mathbbm{G}
\\
&
\frac{1}{2}\, \big[\, \tau_2 + \tau_3 - 1 +  \sqrt{\lambda(1,\, -\, \tau_2,\, \tau_3)} \,\big] - i\, 0^+
&\qquad&
(\tau_2,\, \tau_3) \in \mathbbm{P} \cup \mathbbm{Y}
\\
&
\frac{1}{2}\, \big[\, \tau_2 + \tau_3 - 1 +  \sqrt{\lambda(1,\, -\, \tau_2,\, \tau_3)} \,\big] + i\, 0^+
&\qquad&
(\tau_2,\, \tau_3) \in \mathbbm{R}
\\
&
\vphantom{\frac{1}{1}}{\sqrt{-\, \tau_2}\, e^{\, i \psi}}
&\qquad&
(\tau_2,\, \tau_3) \in \mathbbm{W}
\end{aligned}
\right.
\end{equation}
where
\begin{equation}
\psi = \text{sign} \left( \tau_2^2 - (\tau_3 - 1)^2 \right) \arccos \frac{\tau_2 + \tau_3 - 1}{2\, \sqrt{-\, \tau_2}}\,,
\end{equation}
and the five colored regions $\mathbbm{G}$, $\mathbbm{P}$, $\mathbbm{Y}$, $\mathbbm{R}$ and $\mathbbm{W}$, as visualized in figure \ref{fig6}, are defined as follows:
\begin{equation}
\begin{aligned}
\mathbbm{G} & = \big\{\, (\tau_2,\, \tau_3) \,\big|\, \tau_2 > 0\,, \tau_3 > 0 \,\big\}
\\
\mathbbm{P} & = \big\{\, (\tau_2,\, \tau_3) \,\big|\, \tau_2 > 0\,, \tau_3 < 0 \,\big\}
\,\cup\,
\big\{\, (\tau_2,\, \tau_3) \,\big|\, \sqrt{\tau_3} > \sqrt{-\, \tau_2} + 1 \,\big\}
\\
\mathbbm{Y} & = \big\{\, (\tau_2,\, \tau_3) \,\big|\, \tau_2 < 0\,, \tau_3 < 0 \,\big\}
\,\cup\,
\big\{\, (\tau_2,\, \tau_3) \,\big|\, 1 > \sqrt{-\, \tau_2} + \sqrt{\tau_3} \,\big\}
\\
\mathbbm{R} & = \big\{\, (\tau_2,\, \tau_3) \,\big|\, \sqrt{-\, \tau_2} > \sqrt{\tau_3} + 1 \,\big\}
\\
\mathbbm{W} & = \big\{\, (\tau_2,\, \tau_3) \,\big|\, \sqrt{-\, \tau_2} + \sqrt{\tau_3} > 1 > \left| \sqrt{-\, \tau_2} - \sqrt{\tau_3} \right| \,\big\}
\end{aligned}
\end{equation}
In the multi-colored area, which comprises the green ($\mathbbm{G}$), purple ($\mathbbm{P}$), yellow ($\mathbbm{Y}$) and red ($\mathbbm{R}$) regions, $\xi$ has either a zero or an infinitesimal imaginary part. Conversely, in the white region ($\mathbbm{W}$), $\xi$ is complex with a finite phase. To the right of the bold demarcation in figure \ref{fig6}, i.e., within the green and purple regions, the real part of $\xi$ is positive. On the contrary, in the red and yellow regions located to the left of this line, the real part of $\xi$ is negative. Incorporating the insights from our prior discussion on the analytic continuation of $z$, we can derive the complete analytic continuation for $y$ across the full parameter space. Particularly, for $z > 0$, the analytic continuation of $y$ follows directly from that of $\xi$.

\par
To aid in calculating the $e^+e^- \rightarrow W^+W^-$ process, we summarize the analytic continuations of $x$, $y$ and $z$ to the physical region in table \ref{table:continuation}. The top two rows display the results for the integral family $\mathcal{F}$, as delineated in eq. \eqref{eq:Ffamily}. For the sake of completeness, we also include the results for the integral family $\mathcal{F}^{\ast}$ in the bottom row, which necessitates redefining $x$, $y$ and $z$ by swapping $m_t$ and $m_b$. This table, serving as an essential reference, details the requisite infinitesimal imaginary parts for various kinematic configurations within the physical region.
\begin{table}[htbp]
\centering
\renewcommand\arraystretch{1.4}
\begin{tabular}{
p{1.2cm}<{\centering}
p{4.5cm}<{\centering}
p{1.5cm}<{\centering}
p{1.0cm}<{\centering}
p{1.5cm}<{\centering}
}
\toprule[1.2pt]
\multirow{1}{*}{Family}
&
\multirow{1}{*}{Kinematic configuration}
&
\multirow{1}{*}{$x$}
&
\multirow{1}{*}{$z$}
&
\multirow{1}{*}{$y$}
\\
\midrule[0.4pt]\midrule[0.4pt]
\multirow{2}{*}{$\mathcal{F}$}
&
\multirow{1}{*}{$2\, m_{\scriptscriptstyle{W}} < \sqrt{s} < 2\, m_t$}
&
\multirow{1}{*}{$e^{\, i \vartheta}$}
&
\multirow{1}{*}{$z$}
&
\multirow{1}{*}{$y - i\, 0^+$}
\\
&
\multirow{1}{*}{$2\, m_t < \sqrt{s}$}
&
\multirow{1}{*}{$x + i\, 0^+$}
&
\multirow{1}{*}{$z$}
&
\multirow{1}{*}{$y - i\, 0^+$}
\\
\midrule[0.4pt]
\multirow{1}{*}{~$\mathcal{F}^{\ast}$}
&
\multirow{1}{*}{$2\, m_{\scriptscriptstyle{W}} < \sqrt{s}$}
&
\multirow{1}{*}{$x + i\, 0^+$}
&
\multirow{1}{*}{$z$}
&
\multirow{1}{*}{$y - i\, 0^+$}
\\
\bottomrule[1.2pt]
\end{tabular}
\caption{Analytic continuations of $x$, $y$ and $z$ across various physical kinematic regions.}
\label{table:continuation}
\end{table}

\section{Numerical results and discussion}
\label{sec:4}
\par
Building upon the analytic results for all two-loop canonical MIs presented in section \ref{sec:3}, we calculate the integrated cross section as well as various kinematic distributions of the final-state $W$ bosons for  $e^+e^- \rightarrow W^+W^-$ up to the QCD-EW NNLO. In our calculation, all relevant SM input parameters are set as follows \cite{ParticleDataGroup:2024cfk}:
\begin{align}
\label{eq:SMinput}
& \alpha_s(m_{\scriptscriptstyle{Z}}) = 0.1180\,, &
& \alpha(0)  = 1/137.035999084\,, &
& G_{\mu} = 1.1663788 \times 10^{-5}~ \mathrm{GeV}^{-2}\,,
\nonumber \\
& m_{\scriptscriptstyle{W}} = 80.3692~ \mathrm{GeV}\,, &
& m_{\scriptscriptstyle{Z}} = 91.1880~ \mathrm{GeV}\,, &
& m_{\scriptscriptstyle{H}} = 125.20~ \mathrm{GeV}\,,
\nonumber \\
& m_e = 0.51099895000~ \mathrm{MeV}\,, &
& m_{\mu} = 0.1056583755~ \mathrm{GeV}\,, &
& m_{\tau} = 1.77693~ \mathrm{GeV}\,,
\nonumber \\
& m_u = 0.0530~ \mathrm{GeV}\,, &
& m_c = 1.67~ \mathrm{GeV}\,, &
& m_t = 172.4~ \mathrm{GeV}\,,
\nonumber \\
& m_d = 0.0530~ \mathrm{GeV}\,, &
& m_s = 0.0935~ \mathrm{GeV}\,, &
& m_b = 4.78~ \mathrm{GeV}\,,
\end{align}
where the masses of the $t$, $b$, and $c$ quarks are taken as their pole masses, while the $s$-quark mass is an estimate of the so-called ``current quark mass" in the $\overline{\text{MS}}$ scheme at a renormalization scale of $\mu = 2~ \mathrm{GeV}$. The $u$- and $d$-quark masses, regarded as effective parameters, are adjusted to reproduce the experimentally measured hadronic contribution to the photon vacuum polarization \cite{Denner:1991kt,Jegerlehner:2001ca}:
\begin{equation}
 \Delta \alpha_{\text{had}}^{(5)}(m_{\scriptscriptstyle{Z}})
 =
 0.02783 \pm 0.00006
 =
 \sum_{f=u,d,c,s,b}
\frac{\alpha}{\pi}\, Q_f^2\,
\Big(
\log\frac{m_{\scriptscriptstyle{Z}}^2}{m_f^2} - \frac{5}{3}
\Big)\,.
\end{equation}
We utilize the Mathematica package \texttt{RunDec} \cite{Chetyrkin:2000yt,Herren:2017osy} to evaluate the strong coupling constant $\alpha_s(\mu)$ at $\mu = m_{\scriptscriptstyle{W}}$. Our numerical computations are performed in both the $\alpha(0)$ and $G_{\mu}$ schemes. In the $\alpha(0)$ scheme, the fine structure constant is given in eq. \eqref{eq:SMinput}, while in the $G_{\mu}$ scheme, it is defined as per eq. \eqref{eq:alphaGmu}.

\subsection{Integrated cross sections}
\label{sec:4.2}
\par
\begin{figure}[htbp]
\centering
\includegraphics[width = 1.0\textwidth]{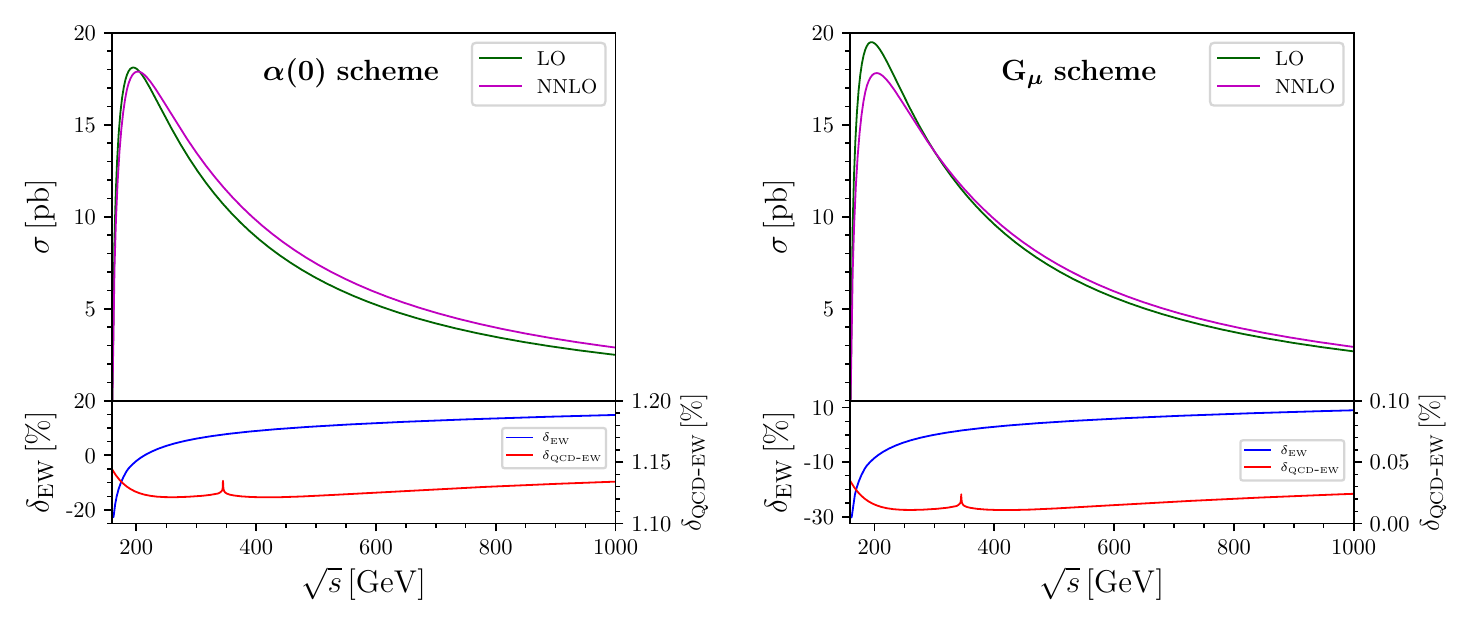}
\caption{LO and NNLO corrected integrated cross sections, along with the corresponding EW and QCD-EW relative corrections, for $e^+e^- \rightarrow W^+W^-$ as functions of the colliding energy.}
\label{fig7}
\end{figure}

The integrated cross section for $e^+e^- \rightarrow W^+W^-$ up to the QCD-EW NNLO can be formulated as
\begin{equation}
\sigma_{\scriptscriptstyle{\text{NNLO}}}
=
\sigma_{\scriptscriptstyle{\text{LO}}}\,
(1 + \delta_{\scriptscriptstyle{\text{EW}}} + \delta_{\scriptscriptstyle{\text{QCD-EW}}})\,,
\end{equation}
where $\delta_{\scriptscriptstyle{\mathrm{EW}}}$ and $\delta_{\scriptscriptstyle{\mathrm{QCD\mbox{-}EW}}}$ represent the NLO EW and NNLO mixed QCD-EW corrections normalized by the LO cross section,
\begin{equation}
\delta_{\scriptscriptstyle{\text{EW}}}
=
\frac{\Delta\sigma_{\scriptscriptstyle{\text{EW}}}}{\sigma_{\scriptscriptstyle{\text{LO}}}}\,,
\qquad\qquad
\delta_{\scriptscriptstyle{\text{QCD-EW}}}
=
\frac{\Delta\sigma_{\scriptscriptstyle{\text{QCD-EW}}}}{\sigma_{\scriptscriptstyle{\text{LO}}}}\,.
\end{equation}
In figure \ref{fig7}, we illustrate the LO and NNLO corrected integrated cross sections as functions of the $e^+e^-$ colliding energy, $\sqrt{s}$, for the process $e^+e^- \rightarrow W^+W^-$ in both the $\alpha(0)$ scheme (left) and the $G_{\mu}$ scheme (right). The corresponding EW and QCD-EW relative corrections are visualized in the lower panels of this figure. The production cross sections show similar trends in both schemes, with a sharp increase near the $W$-pair production threshold and peaking around $\sqrt{s} \sim 195~ \mathrm{GeV}$. Beyond this peak, the cross sections decline smoothly with increasing energy. This behavior is attributed to the interplay between the phase-space expansion and the $s$-channel suppression as the colliding energy increases. In the vicinity of the threshold, the EW corrections significantly reduce the LO cross section by $20-30\%$, transitioning to a moderate increase at higher energies, with enhancements exceeding $14\%$ and $9\%$ at $\sqrt{s} = 1000\, \mathrm{GeV}$ in the $\alpha(0)$ and $G_{\mu}$ schemes, respectively. The remarkable EW corrections near the threshold are due to the Coulomb singularity effect \cite{Denner:1991kt,Beenakker:1996kt}, where the Coulombic photon exchange between the electron and positron significantly enhances the virtual EW corrections as the photon momentum approaches zero. The mixed QCD-EW corrections slightly increase the production cross section across the entire plotted energy region, amounting to approximately $1.1\%$ in the $\alpha(0)$ scheme. However, this effect is considerably attenuated in the $G_{\mu}$ scheme due to the absorption of certain significant higher-order corrections into the LO cross section \cite{Denner:1991kt,Beenakker:1996kt,Denner:2019vbn}. The mixed QCD-EW relative correction peaks at $\sqrt{s} = 2\, m_t$ due to the resonance effect induced by top-quark loop integrals, hitting approximately $1.13\%$ and $0.24\text{\textperthousand}$ at the resonance in the $\alpha(0)$ and $G_{\mu}$ schemes, respectively. The LO, NLO EW and NNLO mixed QCD-EW corrected integrated cross sections, along with the corresponding relative corrections, at some representative colliding energies for both the $\alpha(0)$ and $G_{\mu}$ schemes are summarized in table \ref{table:totalcs}.
\begin{table}[htbp]
\centering
\renewcommand{\arraystretch}{1.4}
\begin{tabular}{
p{1.5cm}<{\centering}
p{1.4cm}<{\centering}
p{1.8cm}<{\centering}
p{1.8cm}<{\centering}
p{1.8cm}<{\centering}
p{1.8cm}<{\centering}
p{1.8cm}<{\centering}
}
\toprule[1.2pt]
\multirow{1}{*}{$\sqrt{s}~ \text{[GeV]}$}
&
\multirow{1}{*}{Scheme}
&
\multirow{1}{*}{$\sigma_{\scriptscriptstyle{\text{LO}}}~ \text{[pb]}$}
&
\multirow{1}{*}{$\sigma_{\scriptscriptstyle{\text{NLO}}}~ \text{[pb]}$}
&
\multirow{1}{*}{$\delta_{\scriptscriptstyle{\text{EW}}}~ \text{[\%]}$}
&
\multirow{1}{*}{$\sigma_{\scriptscriptstyle{\text{NNLO}}}~ \text{[pb]}$}
&
\multirow{1}{*}{$\delta_{\scriptscriptstyle{\text{QCD-EW}}}~ \text{[\%]}$}
\\
\midrule[0.4pt]
\midrule[0.4pt]
\multirow{2}{*}{$161$}
& {$\alpha(0)$}
& {$2.871237$} 
& {$2.24780$}
& {$-21.7131\;\;\,$}
& {$2.28062$}
& {$1.1429$}
\\
& {$G_{\mu}$}
& {$3.089810$}
& {$2.18026$}
& {$-29.4370\;\;\,$}
& {$2.18129$}
& {$0.0333$}
\\
\midrule[0.4pt]                                                                                                 
\multirow{2}{*}{$200$}
& {$\alpha(0)$}
& {$18.04821$}
& {$17.6736$}
& {$\;\,-2.0754\;\;\,$}
& {$17.8768$}
& {$1.1256$}
\\
& {$G_{\mu}$}
& {$19.42213$}
& {$17.7790$}
& {$\;\,-8.4603\;\;\,$}
& {$17.7820$}
& {$0.0154$}
\\
\midrule[0.4pt]                                                                                                  
\multirow{2}{*}{$240$}
& {$\alpha(0)$}
& {$15.93577$}
& {$16.3653$}
& {$\quad\,2.6952\;\;\,$}
& {$16.5440$}
& {$1.1216$}
\\
& {$G_{\mu}$}
& {$17.14888$}
& {$16.5484$}
& {$\;\,-3.5015\;\;\,$}
& {$16.5503$}
& {$0.0113$}
\\
\midrule[0.4pt]                                                                                                 
\multirow{2}{*}{$250$}
& {$\alpha(0)$}
& {$15.31808$}
& {$15.8473$}
& {$\quad\,3.4552\;\;\,$}
& {$16.0191$}
& {$1.1214$}
\\
& {$G_{\mu}$}
& {$16.48417$}
& {$16.0370$}
& {$\;\,-2.7129\;\;\,$}
& {$16.0388$}
& {$0.0110$}
\\
\midrule[0.4pt]                                                                                                 
\multirow{2}{*}{$350$}
& {$\alpha(0)$}
& {$10.49771$}
& {$11.3126$}
& {$\quad\,7.7627\;\;\,$}
& {$11.4307$}
& {$1.1246$}
\\
& {$G_{\mu}$}
& {$11.29685$}
& {$11.4944$}
& {$\quad\,1.7485\;\;\,$}
& {$11.4960$}
& {$0.0143$}
\\
\midrule[0.4pt]                                                                                                 
\multirow{2}{*}{$500$}
& {$\alpha(0)$}
& {$6.673608$}
& {$7.38135$}
& {$\;\;\,10.6051\;\;\,$}
& {$7.45626$}
& {$1.1225$}
\\
& {$G_{\mu}$}
& {$7.181637$}
& {$7.51822$}
& {$\quad\,4.6867\;\;\,$}
& {$7.51909$}
& {$0.0122$}
\\
\midrule[0.4pt]                                                                                             
\multirow{2}{*}{$1000$}
& {$\alpha(0)$}
& {$2.493302$}
& {$2.86231$}
& {$\;\;\,14.8001\;\;\,$}
& {$2.89059$}
& {$1.1341$}
\\
& {$G_{\mu}$}
& {$2.683104$}
& {$2.92488$}
& {$\quad\,9.0110\;\;\,$}
& {$2.92553$}
& {$0.0242$}
\\
\bottomrule[1.2pt]
\end{tabular}
\caption{LO, NLO EW and NNLO QCD-EW corrected integrated cross sections, as well as the corresponding EW and QCD-EW relative corrections, for $e^+e^- \rightarrow W^+W^-$ at some representative colliding energies in both the $\alpha(0)$ and $G_{\mu}$ schemes.}
\label{table:totalcs}
\end{table}

\par
\begin{figure}[htbp]
\centering
\includegraphics[width = 0.93\textwidth]{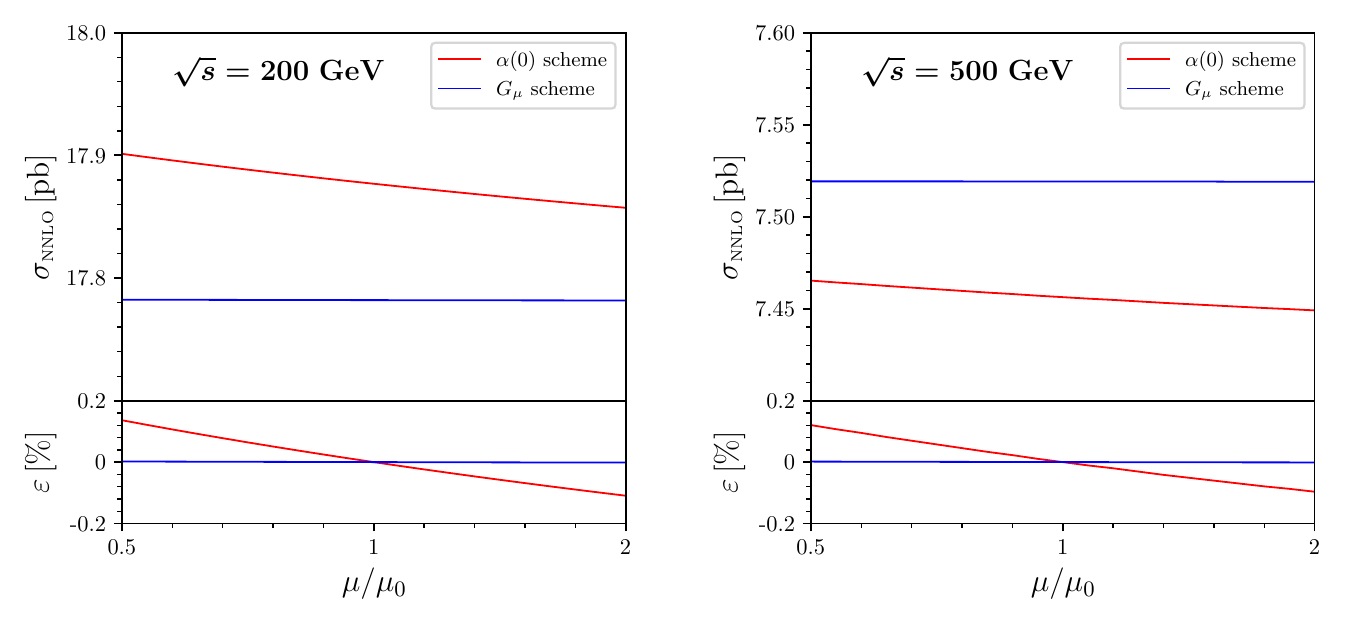}
\caption{Renormalization scale dependence of the NNLO QCD-EW corrected integrated cross sections at $\sqrt{s} = 200$ and $500~ \mathrm{GeV}$ in both the $\alpha(0)$ and $G_{\mu}$ schemes.}
\label{fig8}
\end{figure}

The renormalization scale dependence of the NNLO QCD-EW corrected integrated cross section, arising from the strong coupling constant $\alpha_s(\mu)$, can be directly characterized by
\begin{equation}
\varepsilon(\mu)
=
\frac{\sigma_{\scriptscriptstyle{\text{NNLO}}}(\mu)}{\sigma_{\scriptscriptstyle{\text{NNLO}}}(\mu_0)} - 1\,,
\end{equation}
where the central scale $\mu_0$ is set to $m_{\scriptscriptstyle{W}}$. The variations of $\sigma_{\scriptscriptstyle{\text{NNLO}}}$ and $\varepsilon$ with respect to the renormalization scale $\mu$ over the range $[\mu_0/2,\, 2\mu_0]$ are illustrated in figure \ref{fig8}. It is evident that the scale dependence is quite small, especially in the $G_{\mu}$ scheme. As shown in the lower panels of figure \ref{fig8}, $\varepsilon(\mu)$ decreases monotonically with increasing $\mu$ in the $\alpha(0)$ scheme, ranging from approximately $0.15\%$ to $-0.1\%$ at $\sqrt{s} = 200$ and $500~ \mathrm{GeV}$. In contrast, the variation of $\varepsilon(\mu)$ in the $G_{\mu}$ scheme remains below $0.005\%$, indicating negligible renormalization scale dependence due to the small magnitude of the $\mathcal{O}(\alpha\alpha_s)$ corrections. More detailed and comprehensive numerical results at various colliding energies are presented in table \ref{table:scale}, where the scale uncertainty $\varepsilon_{\text{scale}}$ is defined as
\begin{equation}
\varepsilon_{\text{scale}}
=
\max \left\{
\varepsilon(\mu) - \varepsilon(\mu^{\prime})
\,\big|\,
\mu, \mu^{\prime} \in [\mu_{0}/2,\, 2\mu_0]
\right\}\,.
\end{equation}
In a sense, $\varepsilon_{\text{scale}}$ quantitatively reflects the theoretical error arising from the neglect of higher-order perturbative contributions. Across all colliding energies, the scale uncertainties of $\sigma_{\scriptscriptstyle{\text{NNLO}}}$ are approximately $0.2-0.3\%$ in the $\alpha(0)$ scheme and do not exceed $0.01\%$ in the $G_{\mu}$ scheme, which are roughly of the same order as the NNLO EW corrections. Given the more substantial impact of the mixed QCD-EW corrections in the $\alpha(0)$ scheme, the following discussion will concentrate on the phenomenological analysis within this scheme.
\begin{table}[htbp]
\centering
\renewcommand{\arraystretch}{1.4}
\begin{tabular}{
p{1.5cm}<{\centering}
p{1.4cm}<{\centering}
p{2.0cm}<{\centering}
p{2.0cm}<{\centering}
p{2.0cm}<{\centering}
p{1.8cm}<{\centering}
}
\toprule[1.2pt]
\multirow{1}{*}{$\sqrt{s}~ \text{[GeV]}$}
&
\multirow{1}{*}{Scheme}
&
\multirow{1}{*}{$\sigma(\mu_0/2)~ \text{[pb]} $}
&
\multirow{1}{*}{$\sigma(\mu_0)~ \text{[pb]} $}
&
\multirow{1}{*}{$\sigma(2 \mu_0)~ \text{[pb]} $}
&
\multirow{1}{*}{$\varepsilon_{\text{scale}}~ \text{[\%]}$}
\\
\midrule[0.4pt]
\midrule[0.4pt]
\multirow{2}{*}{$161$}
& \multirow{1}{*}{$\alpha(0)$}
& {$2.28456$}
& {$2.28062$}
& {$2.27745$}
& {$0.31$}
\\
& \multirow{1}{*}{$G_{\mu}$}
& {$2.18141$}
& {$2.18129$}
& {$2.18119$}
&  {$~\,0.010$}
\\
\midrule[0.4pt]
\multirow{2}{*}{$200$}
& \multirow{1}{*}{$\alpha(0)$}
& {$17.9012$}
& {$17.8768$}
& {$17.8572$}
& {$0.25$}
\\
& \multirow{1}{*}{$G_{\mu}$}
& {$17.7823$}
& {$17.7820$}
& {$17.7817$}
&  {$~\,0.003$}
\\
\midrule[0.4pt]
\multirow{2}{*}{$240$}
& \multirow{1}{*}{$\alpha(0)$}
& {$16.5655$}
& {$16.5440$}
& {$16.5268$}
& {$0.23$}
\\
& \multirow{1}{*}{$G_{\mu}$}
& {$16.5506$}
& {$16.5503$}
& {$16.5502$}
&  {$~\,0.002$}
\\
\midrule[0.4pt]
\multirow{2}{*}{$250$}
& \multirow{1}{*}{$\alpha(0)$}
& {$16.0398$}
& {$16.0191$}
& {$16.0026$}
& {$0.23$}
\\
& \multirow{1}{*}{$G_{\mu}$}
& {$16.0390$}
& {$16.0388$}
& {$16.0386$}
&  {$~\,0.002$}
\\
\midrule[0.4pt]
\multirow{2}{*}{$350$}
& \multirow{1}{*}{$\alpha(0)$}
& {$11.4449$}
& {$11.4307$}
& {$11.4193$}
&  {$ 0.22$}
\\
& \multirow{1}{*}{$G_{\mu}$}
& {$11.4962$}
& {$11.4960$}
& {$11.4958$}
& {$~\,0.003$}
\\
\midrule[0.4pt]
\multirow{2}{*}{$500$}
& \multirow{1}{*}{$\alpha(0)$}
& {$7.46526$}
& {$7.45626$}
& {$7.44904$}
& {$0.22$}
\\
& \multirow{1}{*}{$G_{\mu}$}
& {$7.51920$}
& {$7.51909$}
& {$7.51901$}
& {$~\,0.003$}
\\
\midrule[0.4pt]
\multirow{2}{*}{$1000$}
& \multirow{1}{*}{$\alpha(0)$}
& {$2.89399$}
& {$2.89059$}
& {$2.88786$}
& {$0.21$}
\\
& \multirow{1}{*}{$G_{\mu}$}
& {$2.92561$}
& {$2.92553$}
& {$2.92546$}
& {$~\,0.005$}
\\
\bottomrule[1.2pt]
\end{tabular}
\caption{Scale uncertainties of the NNLO QCD-EW corrected integrated cross sections across various colliding energies in both the $\alpha(0)$ and $G_{\mu}$ schemes.}
\label{table:scale}
\end{table}

\subsection{Kinematic distributions}
\label{sec:4.3}
\par
In this subsection, we analyze the scattering angle and transverse momentum distributions of the final-state $W$ bosons for $e^+e^- \rightarrow W^+W^-$. We define the EW and QCD-EW differential relative corrections with respect to the kinematic variable $x$ as
\begin{equation}
\delta_{\scriptscriptstyle{\text{EW}}}
=
\Big(
\frac{\mathrm{d} \sigma_{\scriptscriptstyle{\text{EW}}}}{\mathrm{d} x}
-
\frac{\mathrm{d} \sigma_{\scriptscriptstyle{\text{LO}}}}{\mathrm{d} x}
\Big)
\Big/
\frac{\mathrm{d} \sigma_{\scriptscriptstyle{\text{LO}}}}{\mathrm{d} x}\,,
\qquad
\delta_{\scriptscriptstyle{\text{QCD-EW}}}
=
\Big(
\frac{\mathrm{d} \sigma_{\scriptscriptstyle{\text{QCD-EW}}}}{\mathrm{d} x}
-
\frac{\mathrm{d} \sigma_{\scriptscriptstyle{\text{LO}}}}{\mathrm{d} x}
\Big)
\Big/
\frac{\mathrm{d} \sigma_{\scriptscriptstyle{\text{LO}}}}{\mathrm{d} x}\,.
\end{equation}
Due to $\mathcal{CP}$ conservation, the differential distributions of $W$ bosons exhibit the following symmetry relations:
\begin{equation}
\frac{\mathrm{d} \sigma}{\mathrm{d} \cos\theta_{\scriptscriptstyle{W^-}}}
=
\frac{\mathrm{d} \sigma}{\mathrm{d} \cos\theta_{\scriptscriptstyle{W^+}}}\Big|_{\theta \rightarrow \pi - \theta}\,,
\qquad\qquad
\frac{\mathrm{d} \sigma}{\mathrm{d} p_{T,\scriptscriptstyle{W^-}}}
=
\frac{\mathrm{d} \sigma}{\mathrm{d} p_{T,\scriptscriptstyle{W^+}}}
\end{equation}
Consequently, our subsequent discussion will focus solely on the differential distributions of $W^+$, with ${\mathrm{d} \sigma}/{\mathrm{d} \cos\theta}$ representing the scattering angle distribution and ${\mathrm{d} \sigma}/{\mathrm{d} p_T}$ representing the transverse momentum distribution of the $W^+$ boson.

\par
\begin{figure}[htbp]
\centering
\includegraphics[width = 1.0\textwidth]{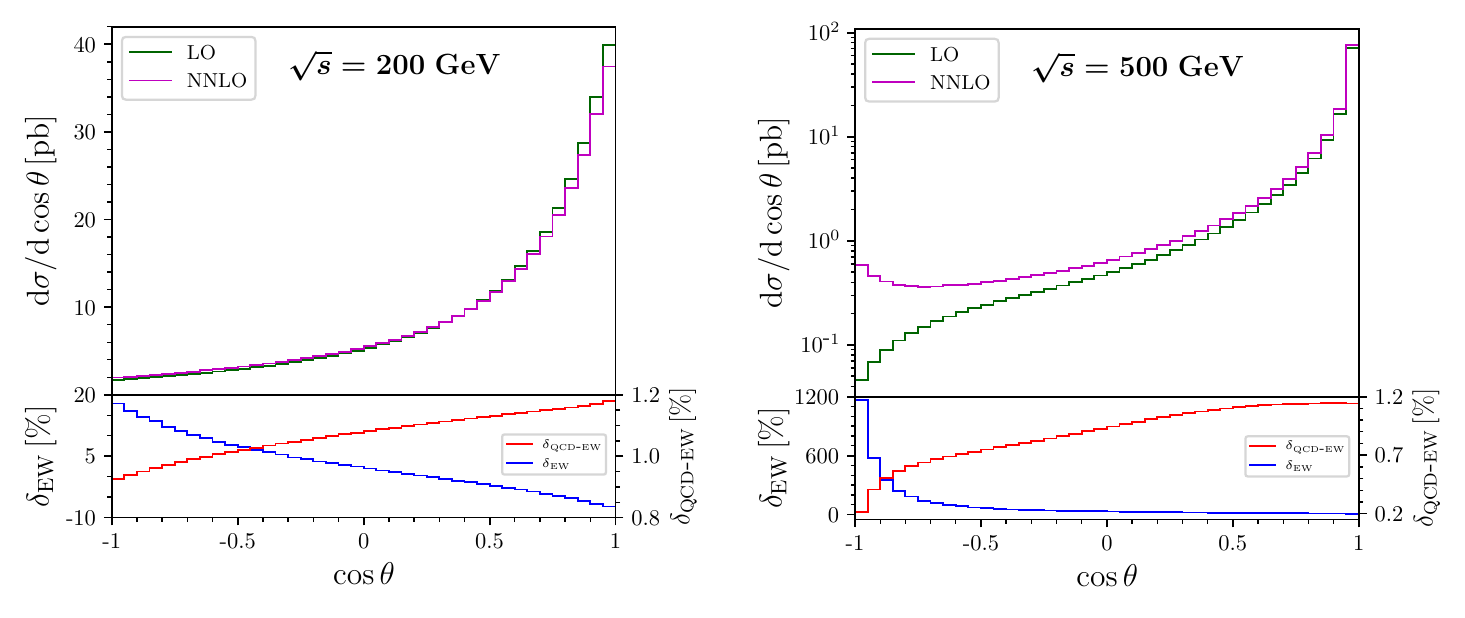}
\caption{LO and NNLO corrected scattering angle distributions of the final-state $W^+$ boson, along with the corresponding EW and QCD-EW relative corrections for $e^+e^- \rightarrow W^+W^-$, at $\sqrt{s}= 200$ (left) and $500~ \mathrm{GeV}$ (right).}
\label{fig9}
\end{figure}

In figure \ref{fig9}, we present the LO and NNLO corrected scattering angle distributions of the final-state $W^+$ boson, along with the corresponding EW and QCD-EW relative corrections, for $e^+e^- \rightarrow W^+W^-$ at CM energies of $\sqrt{s} = 200$ and $500~ \mathrm{GeV}$. The scattering angle distribution is notably peaked in the forward direction, especially at high energies, and diminishes progressively with increasing scattering angle. At $\sqrt{s} = 200\, \mathrm{GeV}$, the EW correction exhibits a moderate enhancement of approximately $18\%$ to the LO differential cross section in the backward direction, transitioning into a suppression of around $-7\%$ in the forward direction. This increase in the backward direction is attributed to the boost effect caused by hard photon emissions, which propels the CM system of the $W$-boson pair. This effect leads to a migration of contributions from the forward region to the backward region, and vice versa. Given that the forward cross section is significantly greater than that of the backward region, the resultant redistribution distorts the scattering angle distribution relative to the LO distribution. This boost effect becomes more pronounced at high energies, as evidenced by the comparison of the EW relative corrections at $\sqrt{s} = 200$ and $500~ \mathrm{GeV}$ depicted in the two lower panels of figure \ref{fig9}. At extreme backward angles, the NLO EW correction can exceed the LO cross section by an order of magnitude when $\sqrt{s} > 500~ \mathrm{GeV}$, challenging the perturbative reliability in this kinematic region.\footnote{For further details and treatments on the boost effect, please refer to refs. \cite{Zerwas:1991rrh,Beenakker:1994vn,Beenakker:1996kt}} In contrast to the NLO EW corrections, the NNLO mixed QCD-EW corrections to $e^+e^- \rightarrow W^+W^-$ do not exhibit the boost effect, due to the absence of photon and gluon emissions at this order. The NNLO QCD-EW corrections slightly enhance the LO scattering angle distribution across the entire range of $\cos\theta$. The QCD-EW relative correction is more sensitive to $\cos\theta$ in the backward region compared to the forward region, increasing monotonically with $\cos\theta$. At $\sqrt{s} = 200~ \mathrm{GeV}$, the QCD-EW relative correction increases from approximately $0.9\%$ to about $1.2\%$ as $\cos\theta$ moves from $-1$ to $1$. At a higher energy of $\sqrt{s} = 500~ \mathrm{GeV}$, it varies from around $0.2\%$ to about $1.1\%$ with the increase of $\cos\theta$ from $-1$ to $0.4$, and then levels off in the rest range of $\cos\theta$, stabilizing at around $1.1\%$.

\par
\begin{figure}[htbp]
\centering
\includegraphics[width = 1.0\textwidth]{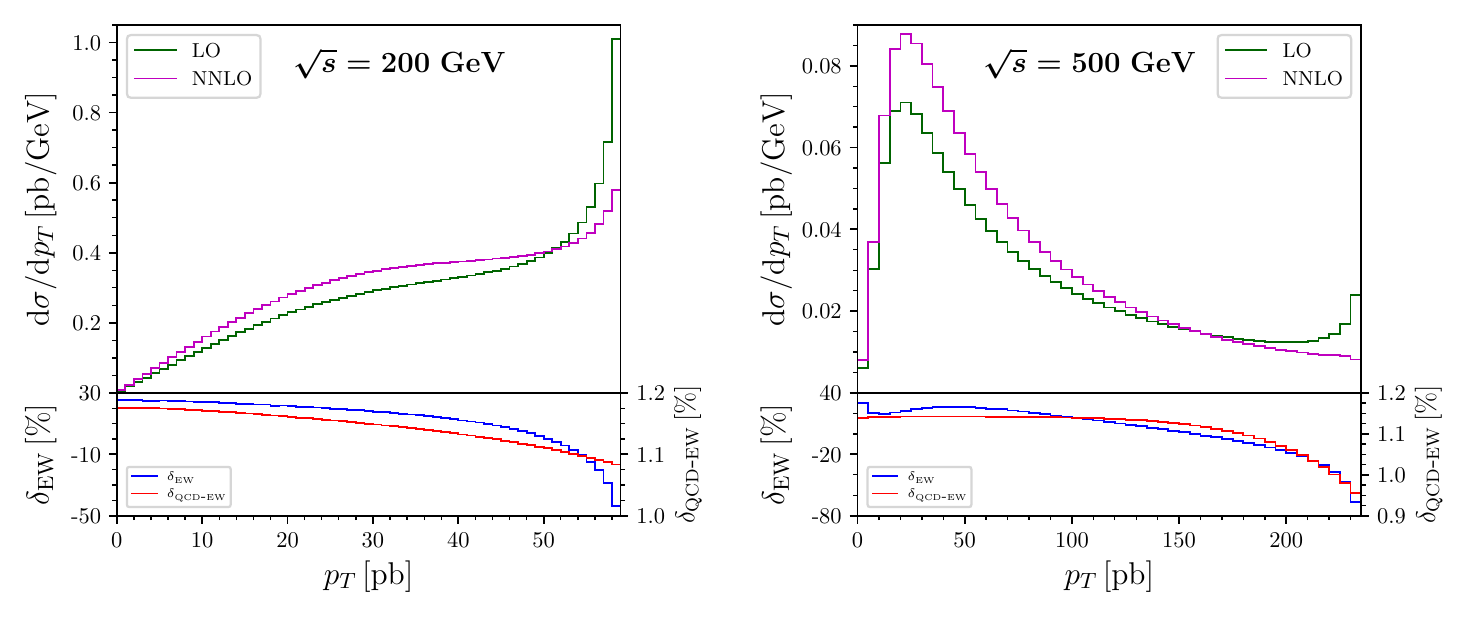}
\caption{Same as figure \ref{fig9}, but for transverse momentum distributions of the final-state $W^+$ boson.}
\label{fig10}
\end{figure}

The LO and NNLO corrected transverse momentum distributions of the final-state $W^+$ boson, along with the corresponding EW and QCD-EW relative corrections, are plotted in figure \ref{fig10}. As can be seen from this figure, the $W$-boson pairs are predominantly produced in the high $p_T$ region at $\sqrt{s} = 200~ \mathrm{GeV}$, while the events are mostly concentrated in the low $p_T$ region at $\sqrt{s} = 500~\mathrm{GeV}$. The NLO EW corrections enhance the LO differential cross section in the low $p_T$ region and suppress it in the high $p_T$ region. The pronounced magnitude of the EW relative correction at extremely high $p_T$ can be largely attributed to the Sudakov effect. The QCD-EW relative correction exhibits increased sensitivity to $p_T$ as $p_T$ increases, particularly at high colliding energies. At $\sqrt{s} = 500~\mathrm{GeV}$, this correction remains relatively constant at roughly $1.1\%$ for $p_T < 150~ \mathrm{GeV}$ and decreases rapidly for higher $p_T$. By considering both $\cos\theta$ and $p_T$ distributions, it is apparent that the mixed QCD-EW correction exceeds $0.9\%$ across most of the phase space, with the potential to reach approximately $1.2\%$. This indicates that the NNLO mixed QCD-EW corrections are non-negligible and should be taken into account when comparing theoretical predictions with future high-precision experimental data, especially in certain phase-space regions.

\section{Summary}
\label{sec:5}
The recent discrepancy between the CDF measurement and the SM prediction for the $W$-boson mass highlights the need for further refinement in both theoretical predictions and experimental measurements. In this work, we detail a comprehensive calculation of the NNLO mixed QCD-EW corrections to $W$-pair production at electron-positron colliders. By employing the canonical differential equation method, we analytically calculate all two-loop MIs necessary for these corrections, obtaining $32$ triangle canonical MIs expressed in terms of GPLs up to the order of $\epsilon^4$. Using these analytic expressions, we compute the total production cross section and the differential distributions with respect to the scattering angle and transverse momentum of the final-state $W^+$ boson in both the $\alpha(0)$ and $G_{\mu}$ schemes. Our findings indicate that the NNLO mixed QCD-EW corrections enhance the LO cross section. Specifically, in the $\alpha(0)$ scheme, the relative correction exceeds $0.9\%$ across most of the phase space and can approach $1.2\%$ in certain phase-space regions, and thus is non-negligible for comparisons with future high-precision experimental measurements. In contrast, the NNLO QCD-EW correction is relatively minor in the $G_{\mu}$ scheme. To further reduce the dependence of theoretical predictions on the chosen scheme for $\alpha$, it is essential to also consider the NNLO pure EW $\mathcal{O}(\alpha^2)$ corrections, which are beyond the scope of this paper and will be addressed in future research.

\acknowledgments
\par
This work is supported by the National Natural Science Foundation of China (Grant No. 12061141005) and the CAS Center for Excellence in Particle Physics (CCEPP).

\appendix
\section{Explicit expressions of canonical MIs}
\label{appendix:A}
\par
The explicit analytic expressions of the 32 canonical MIs $\text{I}_{i}$ ($i = 1, \dots, 32$) in terms of GPLs up to $\mathcal{O}(\epsilon^{2})$ are listed as follows:
{
\fontsize{9pt}{15}\selectfont
\begin{align}
\text{I}_{1} = {} &
                    1
\nonumber \\
\text{I}_{2} = {} &
                    1
                    - \epsilon\,
                    \big[\,
                    G(-1; y)
                    - 2\, G(-1; z)
                    + G(-y-1; z)
                    + G({xy}/{((1-x)^{2}-xy)}; z)
                    \,\big]
\nonumber \\
&
                    - \epsilon^{2}\,
                    \big\{\,
                    G(-1; y)\,
                    \big[\,
                    2\, G(-1; z)
                    - G(-y-1; z)
                    - G({xy}/{((1-x)^{2}-xy)}; z)
                    \,\big]
                    - G(-1, -1; y)
\nonumber \\
&
                    - 4\, G(-1, -1; z)
                    - G(-y-1, -y-1; z)
                    - G({xy}/{((1-x)^{2}-xy)}, {xy}/{((1-x)^{2}-xy)}; z)
\nonumber \\
&
                    + 2\, G(-y-1, -1; z)
                    + 2\, G({xy}/{((1-x)^{2}-xy)}, -1; z)
                    - G({xy}/{((1-x)^{2}-xy)}, -y-1; z)
\nonumber \\
&
                    + 2\, G(-1, -y-1; z)
                    + 2\, G(-1, {xy}/{((1-x)^{2}-xy)}; z)
                    - G(-y-1, {xy}/{((1-x)^{2}-xy)}; z)
                    \,\big\}
\nonumber \\
\text{I}_{3} = {} &
                    1
                    - 2\, \epsilon\,
                    \big[\,
                    G(-1; y)
                    - 2\, G(-1; z)
                    + G(-y-1; z)
                    + G({xy}/{((1-x)^{2}-xy)}; z)
                    \,\big]
\nonumber \\
&
                    - 4\, \epsilon^{2}\,
                    \big\{\,
                    G(-1; y)\,
                    \big[\,
                    2\, G(-1; z)
                    - G(-y-1; z)
                    - G({xy}/{((1-x)^{2}-xy)}; z)
                    \,\big]
                    - G(-1, -1; y)
\nonumber \\
&
                    - 4\, G(-1, -1; z)
                    - G(-y-1, -y-1; z)
                    - G({xy}/{((1-x)^{2}-xy)}, {xy}/{((1-x)^{2}-xy)}; z)
\nonumber \\
&
                    + 2\, G(-y-1, -1; z)
                    + 2\, G({xy}/{((1-x)^{2}-xy)}, -1; z)
                    - G({xy}/{((1-x)^{2}-xy)}, -y-1; z)
\nonumber \\
&
                    + 2\, G(-1, -y-1; z)
                    + 2\, G(-1, {xy}/{((1-x)^{2}-xy)}; z)
                    - G(-y-1, {xy}/{((1-x)^{2}-xy)}; z)
                    \,\big\}
\nonumber \\
\text{I}_{4} = {} &
                    \epsilon\, G(0; x)
                    + \epsilon^{2}\,
                    \big[\,
                    G(0, 0; x)
                    - 2\, G(-1, 0; x)
                    - \pi^{2}/6
                    \,\big]
\nonumber \\
\text{I}_{5} = {} &
                    \epsilon\, G(0; x)
                    - \epsilon^{2}\,
                    \big\{\,
                    G(0; x)\,
                    \big[\,
                    G(-1; y)
                    - 2\, G(-1; z)
                    + G(-y-1; z)
                    + G({xy}/{((1-x)^{2}-xy)}; z)
                    \,\big]
\nonumber \\
&
                    - G(0, 0; x)
                    + 2\, G(-1, 0; x)
                    + \pi^{2}/6
                    \,\big\}
\nonumber \\
\text{I}_{6} = {} &
                    - 1/2\, \epsilon\,
                    \big[\,
                    G(-1; y)
                    + G(-y-1; z)
                    - G({xy}/{((1-x)^{2}-xy)}; z)
                    \,\big]
\nonumber \\
&
                    + 1/2\, \epsilon^{2}\,
                    \big\{\,
                    G(-1; y)\,
                    \big[\,
                    G(-y-1; z)
                    + 2\, G(-{xy^{2}}/{(1-x)^{2}}; z)
                    - G({xy}/{((1-x)^{2}-xy)}; z)
                    \,\big]
\nonumber \\
&
                    - G({xy}/{((1-x)^{2}-xy)}, -y-1; z)
                    - 2\, G(0, -1; z)
                    + 2\, G({xy}/{((1-x)^{2}-xy)}, -1; z)
\nonumber \\
&
                    + G(-y-1, {xy}/{((1-x)^{2}-xy)}; z)
                    + G(-1, -1; y)
                    + 2\, G(-{xy^{2}}/{(1-x)^{2}}, -y-1; z)
\nonumber \\
&
                    - 2\, G(-y-1, -1; z)
                    + G(-y-1, -y-1; z)
                    - G({xy}/{((1-x)^{2}-xy)}, {xy}/{((1-x)^{2}-xy)}; z)
\nonumber \\
&
                    + 2\, G(0, {xy}/{((1-x)^{2}-xy)}; z)
                    - 2\, G(-{xy^{2}}/{(1-x)^{2}}, {xy}/{((1-x)^{2}-xy)}; z)
                    \,\big\}
\nonumber \\
\text{I}_{7} = {} &
                    - 1/2\, \epsilon\,
                    \big[\,
                    G(-1; y)
                    + G(-y-1; z)
                    - G({xy}/{((1-x)^{2}-xy)}; z)
                    \,\big]
\nonumber \\
&
                    - 1/2\, \epsilon^{2}\,
                    \big\{\,
                    G(-1; y)\,
                    \big[\,
                    2\, G(-1; z)
                    - 3\, G(-y-1; z)
                    - 2\, G(-{xy^{2}}/{(1-x)^{2}}; z)
                    + G({xy}/{((1-x)^{2}-xy)}; z)
                    \,\big]
\nonumber \\
&
                    + 4\, G(-y-1, -1; z)
                    + G({xy}/{((1-x)^{2}-xy)}, -y-1; z)
                    - 4\, G({xy}/{((1-x)^{2}-xy)}, -1; z)
\nonumber \\
&
                    + 2\, G(-1, -y-1; z)
                    - G(-y-1, {xy}/{((1-x)^{2}-xy)}; z)
                    - 2\, G(-1, {xy}/{((1-x)^{2}-xy)}; z)
\nonumber \\
&
                    - 3\, G(-1, -1; y)
                    - 3\, G(-y-1, -y-1; z)
                    + 3\, G({xy}/{((1-x)^{2}-xy)}, {xy}/{((1-x)^{2}-xy)}; z)
\nonumber \\
&
                    + 2\, G(0, -1; z)
                    - 2\, G(0, {xy}/{((1-x)^{2}-xy)}; z)
                    - 2\, G(-{xy^{2}}/{(1-x)^{2}}, -y-1; z)
\nonumber \\
&
                    + 2\, G(-{xy^{2}}/{(1-x)^{2}}, {xy}/{((1-x)^{2}-xy)}; z)
                    \,\big\}
\nonumber \\
\text{I}_{8} = {} &
                    2\, \epsilon^{2}\, G(0, 0; x)
\nonumber \\
\text{I}_{9} = {} &
                    - 1/2\, \epsilon^{2}\, G(0; x)\,
                    \big[\,
                    G(-1; y)
                    + G(-y-1; z)
                    - G({xy}/{((1-x)^{2}-xy)}; z)
                    \,\big]
\nonumber \\
\text{I}_{10} = {} &
                    1/2\, \epsilon^{2}\,
                    \big\{\,
                    G(-1; y)\,
                    \big[\,
                    G(-y-1; z)
                    - G({xy}/{((1-x)^{2}-xy)}; z)
                    \,\big]
                    + G(-1, -1; y)
\nonumber \\
&
                    - G({xy}/{((1-x)^{2}-xy)}, -y-1; z)
                    - G(-y-1, {xy}/{((1-x)^{2}-xy)}; z)
\nonumber \\
&
                    + G(-y-1, -y-1; z)
                    + G({xy}/{((1-x)^{2}-xy)}, {xy}/{((1-x)^{2}-xy)}; z)
                    \,\big\}
\nonumber \\
\text{I}_{11} = {} &
                    - \epsilon^{2}\,
                    \big\{\,
                    G(0; x)\,
                    \big[\,
                    G(x-1; y)
                    - G({(1-x)}/{x}; y)
                    - G({y}/{(x-1)}; z)
                    + G({xy}/{(1-x)}; z)
                    \,\big]
\nonumber \\
&
                    + G(-1; y)\,
                    \big[\,
                    G({y}/{(x-1)}; z)
                    + G({xy}/{(1-x)}; z)
                    \,\big]
                    + G(0, 0; x)
                    - G({(1-x)}/{x}, -1; y)
\nonumber \\
&
                    - G(x-1, -1; y)
                    + 2\, G(0, -1; y)
                    - 2\, G(0, -1; z)
                    + 2\, G(0, {xy}/{((1-x)^{2}-xy)}; z)
\nonumber \\
&
                    + G({y}/{(x-1)}, -y-1; z)
                    - G({xy}/{(1-x)}, {xy}/{((1-x)^{2}-xy)}; z)
\nonumber \\
&
                    + G({xy}/{(1-x)}, -y-1; z)
                    - G({y}/{(x-1)}, {xy}/{((1-x)^{2}-xy)}; z)
                    \,\big\}
\nonumber \\
\text{I}_{12} = {} &
                    \text{I}_{11}
\nonumber \\
\text{I}_{13} = {} &
                    0
\nonumber \\
\text{I}_{14} = {} &
                    0
\nonumber \\
\text{I}_{15} = {} &
                    \text{I}_{8}
\nonumber \\
\text{I}_{16} = {} &
                    - \epsilon\, G(0; x)
                    - \epsilon^{2}\,
                    \big[\,
                    4\, G(0, 0; x)
                    - 2\, G(1, 0; x)
                    - 6\, G(-1, 0; x)
                    - \pi^{2}/6
                    \,\big]
\nonumber \\
\text{I}_{17} = {} &
                    \epsilon^{2}\,
                    \big\{\,
                    G(-1; y)\,
                    \big[\,
                    G(-1; z)
                    - G({xy}/{((1-x)^{2}-xy)}; z)
                    \,\big]
                    - 2\, G(-1, -1; z)
\nonumber \\
&
                    + G(-y-1, -1; z)
                    + G({xy}/{((1-x)^{2}-xy)}, -1; z)
                    - G({xy}/{((1-x)^{2}-xy)}, -y-1; z)
\nonumber \\
&
                    + G(-1, -y-1; z)
                    + G(-1, {xy}/{((1-x)^{2}-xy)}; z)
                    - G(-y-1, {xy}/{((1-x)^{2}-xy)}; z)
                    \,\big\}
\nonumber \\
\text{I}_{18} = {} &
                    \epsilon\,
                    \big[\,
                    G(-1; y)
                    + G(-y-1; z)
                    - G({xy}/{((1-x)^{2}-xy)}; z)
                    \,\big]
\nonumber \\
&
                    - \epsilon^{2}\,
                    \big\{\,
                    G(-1; y)\,
                    \big[\,
                    G(-y-1; z)
                    + 6\, G(-{xy^{2}}/{(1-x)^{2}}; z)
                    - 3\, G({xy}/{((1-x)^{2}-xy)}; z)
                    \,\big]
\nonumber \\
&
                    + 2\, G(0, -1; y)
                    - 3\, G({xy}/{((1-x)^{2}-xy)}, -y-1; z)
                    + 4\, G({xy}/{((1-x)^{2}-xy)}, -1; z)
\nonumber \\
&
                    - 4\, G(0, -1; z)
                    + 3\, G(-y-1, {xy}/{((1-x)^{2}-xy)}; z)
                    + 6\, G(-{xy^{2}}/{(1-x)^{2}}, -y-1; z)
\nonumber \\
&
                    + G(-1, -1; y)
                    + G(-y-1, -y-1; z)
                    - G({xy}/{((1-x)^{2}-xy)}, {xy}/{((1-x)^{2}-xy)}; z)
\nonumber \\
&
                    - 4\, G(-y-1, -1; z)
                    + 4\, G(0, {xy}/{((1-x)^{2}-xy)}; z)
                    - 6\, G(-{xy^{2}}/{(1-x)^{2}}, {xy}/{((1-x)^{2}-xy)}; z)
                    \,\big\}
\nonumber \\
\text{I}_{19} = {} &
                    - 1/2\, \epsilon\,
                    \big[\,
                    G(-1; y)
                    - 2\, G(-1; z)
                    + G(-y-1; z)
                    + G({xy}/{((1-x)^{2}-xy)}; z)
                    \,\big]
\nonumber \\
&
                    - 1/4\, \epsilon^{2}\,
                    \big\{\,
                    G(-1; y)\,
                    \big[\,
                    11\, G(-1; z)
                    - 2\, G(-y-1; z)
                    - 5\, G({xy}/{((1-x)^{2}-xy)}; z)
                    \,\big]
\nonumber \\
&
                    - 2\, G(-1, -1; y)
                    - 4\, G(0, -1; y)
                    + 4\, G(0, -1; z)
                    - 4\, G(0, {xy}/{((1-x)^{2}-xy)}; z)
\nonumber \\
&
                    - 5\, G({xy}/{((1-x)^{2}-xy)}, -y-1; z)
                    + 11\, G(-1, {xy}/{((1-x)^{2}-xy)}; z)
                    + 7\, G(-y-1, -1; z)
\nonumber \\
&
                    - 5\, G(-y-1, {xy}/{((1-x)^{2}-xy)}; z)
                    + 7\, G({xy}/{((1-x)^{2}-xy)}, -1; z)
                    + 11\, G(-1, -y-1; z)
\nonumber \\
&
                    - 22\, G(-1, -1; z)
                    - 2\, G(-y-1, -y-1; z)
                    - 2\, G({xy}/{((1-x)^{2}-xy)}, {xy}/{((1-x)^{2}-xy)}; z)
                    \,\big\}
\nonumber \\
\text{I}_{20} = {} &
                    0
\nonumber \\
\text{I}_{21} = {} &
                    - 1/2\, \text{I}_{11}
\nonumber \\
\text{I}_{22} = {} &
                    1/2\, \epsilon^{2}\,
                    \big\{\,
                    G(-1; y)\,
                    \big[\,
                    2\, G(0; x)
                    + G({y}/{(x-1)}; z)
                    - G({xy}/{(1-x)}; z)
                    \,\big]
                    - G(0; x)\,
                    \big[\,
                    G({xy}/{(1-x)}; z)
\nonumber \\
&
                    + G({y}/{(x-1)}; z)
                    - 2\, G(-y-1; z)
                    + 2\, G(-1; z)
                    - 2\, G({xy}/{((1-x)^{2}-xy)}; z)
                    + G({(1-x)}/{x}; y)
\nonumber \\
&
                    + G(x-1; y)
                    \,\big]
                    + G(0, 0; x)
                    - 2\, G(1, 0; x)
                    - G({xy}/{(1-x)}, -y-1; z)
                    + G({y}/{(x-1)}, -y-1; z)
\nonumber \\
&
                    - G({y}/{(x-1)}, {xy}/{((1-x)^{2}-xy)}; z)
                    + G({xy}/{(1-x)}, {xy}/{((1-x)^{2}-xy)}; z)
                    + G(x-1, -1; y)
\nonumber \\
&
                    - G({(1-x)}/{x}, -1; y)
                    + \pi^{2}/3
                    \,\big\}
\nonumber \\
\text{I}_{23} = {} &
                    0
\nonumber \\
\text{I}_{24} = {} &
                    - 1/2\, \text{I}_{11}
\nonumber \\
\text{I}_{25} = {} &
                    1/8\, \epsilon\,
                    \big[\,
                    G(-1; y)
                    + G(-y-1; z)
                    - G({xy}/{((1-x)^{2}-xy)}; z)
                    \,\big]
\nonumber \\
&
                    - 1/8\, \epsilon^{2}\,
                    \big\{\,
                    4\, G(0; x)\,
                    \big[\,
                    G(x-1; y)
                    - G({(1-x)}/{x}; y)
                    + G({y}/{(x-1)}; z)
                    - G({xy}/{(1-x)}; z)
                    \,\big]
\nonumber \\
&
                    + G(-1; y)\,
                    \big[\,
                    G(-y-1; z)
                    - 4\, G({y}/{(x-1)}; z)
                    - 4\, G({xy}/{(1-x)}; z)
                    + 6\, G(-{xy^{2}}/{(1-x)^{2}}; z)
\nonumber \\
&
                    + G({xy}/{((1-x)^{2}-xy)}; z)
                    \,\big]
                    + 6\, G(0, -1; y)
                    - 4\, G(x-1, -1; y)
                    - 4\, G({(1-x)}/{x}, -1; y)
\nonumber \\
&
                    + G(-1, -1; y)
                    + 6\, G(-{xy^{2}}/{(1-x)^{2}}, -y-1; z)
                    - 6\, G(-{xy^{2}}/{(1-x)^{2}}, {xy}/{((1-x)^{2}-xy)}; z)
\nonumber \\
&
                    + G(-y-1, -y-1; z)
                    - G({xy}/{((1-x)^{2}-xy)}, {xy}/{((1-x)^{2}-xy)}; z)
\nonumber \\
&
                    - G(-y-1, {xy}/{((1-x)^{2}-xy)}; z)
                    + G({xy}/{((1-x)^{2}-xy)}, -y-1; z)
\nonumber \\
&
                    + 4\, G({xy}/{(1-x)}, {xy}/{((1-x)^{2}-xy)}; z)
                    - 4\, G({y}/{(x-1)}, -y-1; z)
\nonumber \\
&
                    + 4\, G({y}/{(x-1)}, {xy}/{((1-x)^{2}-xy)}; z)
                    - 4\, G({xy}/{(1-x)}, -y-1; z)
                    \,\big\}
\nonumber \\
\text{I}_{26} = {} &
                    0
\nonumber \\
\text{I}_{27} = {} &
                    0
\nonumber \\
\text{I}_{28} = {} &
                    0
\nonumber \\
\text{I}_{29} = {} &
                    - \text{I}_{8}
\nonumber \\
\text{I}_{30} = {} &
                    0
\nonumber \\
\text{I}_{31} = {} &
                    0
\nonumber \\
\text{I}_{32} = {} &
                    \text{I}_{8}
\nonumber
\end{align}
}

\bibliographystyle{JHEP}
\bibliography{references}

\end{document}